\documentclass[utf8]{article}

\usepackage{url,hyperref,lineno,microtype,subcaption}
\usepackage[onehalfspacing]{setspace}
\usepackage{verbatim}
\usepackage{placeins}
\usepackage{booktabs}
\usepackage{ulem}
\usepackage{upgreek}
\usepackage{newtxmath}
\usepackage{graphicx}

\newcommand{\ie}{i.\,e.\ {}}
\newcommand{\iec}{i.\,e., {}}
\newcommand{\eg}{e.\,g.\ {}}
\newcommand{\egc}{e.\,g., {}}

\begin{document}
\onecolumn

\title{Towards realistic HPC models of the neuromuscular system} 

\author{Chris Bradley$^6$, Nehzat Emamy $^{3,5}$, Thomas Ertl $^{4,5}$, Dominik G\"oddeke $^{2,5}$, \\ Andreas Hessenthaler $^{1,5}$, 
Thomas Klotz $^{1,5}$, Aaron Kr\"amer $^{2,5}$, Michael Krone $^{4,5}$, \\ Benjamin Maier $^{3,5}$, Miriam Mehl $^{3,5}$,
Tobias Rau $^{4,5}$, Oliver R\"ohrle \,$^{1,5, }$}

\maketitle

\noindent
$^{1}$Institute of Applied Mechanics (CE), SimTech Research Group on Continuum Biomechanics \\ and Mechanobiology, University of Stuttgart \\
$^{2}$Institute for Applied Analysis and Numerical Simulation, University of Stuttgart \\
$^{3}$Institute for Parallel and Distributed Systems, University of Stuttgart \\
$^{4}$Visualization Research Center of the University of Stuttgart (VISUS), University of Stuttgart \\
$^{5}$Stuttgart Centre for Simulation Sciences, University of Stuttgart, Germany \\
$^{6}$Auckland Bioengineering Institute, The University of Auckland, New Zealand

\begin{abstract}


Realistic simulations of detailed, biophysics-based, multi-scale models require very high resolution and, thus, large-scale compute facilities. 
Existing simulation environments, especially for biomedical applications, are designed to allow for a high flexibility and generality in model development. Flexibility and model development, however, are often a limiting factor for large-scale simulations. Therefore,  new models are typically tested and run on small-scale compute facilities. 
By using a detailed biophysics-based, chemo-electromechanical skeletal muscle model and the international open-source software library OpenCMISS as an example, we present an approach to upgrade an existing muscle simulation framework from a moderately parallel version towards a massively parallel one that scales both in terms of problem size and in terms of the number of parallel processes.
For this purpose, we investigate different modeling, algorithmic and implementational aspects.
We present improvements addressing both numerical and parallel scalability. In addition, our approach includes a novel visualization environment, which is based on the MegaMol environment capable of handling large amounts of simulated data. It offers a platform for fast visualization prototyping, distributed rendering, and advanced visualization techniques.
We present results of a variety of scaling studies at the Tier-1 supercomputer HazelHen at the High Performance Computing Center Stuttgart (HLRS). 
We improve the overall runtime by a factor of up to 2.6 and achieved good scalability on up to 768 cores, where the previous implementation  used only 4 cores. \\
 
\end{abstract}

\section{Introduction}

Even ``simple'' tasks like grabbing an object involve highly coordinated actions of our musculoskeletal system. At the core of such coordinated movements are voluntary contractions of skeletal muscles. Understanding the underlying mechanism of recruitment and muscle force generation is a challenging task and subject to much research. One of the few non-invasive and clinically available diagnostic tools to obtain insights in the functioning (or disfunctioning) of the neuromuscular system are electromyographic (EMG) recordings, e.g., measuring the activation-induced, resulting potentials on the skin surface. Conclusions on the neuromuscular system then reduce to a signal processing task, which often is based on the assumption that the muscle itself is a black box and hence ignores spatial arrangements within the respective muscle. But (surface) EMG measurements also have their limitations, e.g., they typically only capture activity from muscle parts close to the surface, which leads to difficulties in identifying cross-talk. In addition, it often only records weak signals due to layers of adipose tissue, or is, in some cases, restricted to isometric contractions. Hence, to obtain more holistic and non-invasive insights into the neuromuscular system, computational models can be employed. Such models need to capture much of the electro-mechanical properties of skeletal muscle tissue and of the interaction between neural recruitment and muscular contraction. 

The contractile behavior of skeletal muscle tissue has been extensively modeled using lumped-parameter models such as Hill-type skeletal muscle models \cite{Zajac1989}, continuum-mechanical skeletal muscle models \cite{Johansson2000, Blemker2005a, Roehrle2007, Boel2008}, or multi-scale, chemo-electromechanical skeletal muscle models \cite{Roehrle2008, Roehrle2012, HernandezGascon2013, Heidlauf2013}. For prediction of the resulting EMG of a particular stimulation, there exist analytical models \cite{Dimitrov1998, Farina2001, Mesin2006} with short compute times, or numerical approaches \cite{Lowery2002, Mesin2006, Mordhorst2015, Mordhorst2017}. 

However, irrespective of whether one aims to simulate EMG signals or the mechanical behavior of a skeletal muscle, numerical methods are almost unavoidable, if realistic muscle geometries are considered. The chemo-electromechanical models as proposed by \cite{Roehrle2012}, \cite{Heidlauf2013,Heidlauf2014}, or \cite{Heidlauf2016} are particularly well-suited to incorporate many structural and functional features of skeletal muscles. They embed one-dimensional ``muscle fibers'' within a three-dimensional skeletal muscle model and associate them with a particular motor unit. Moreover, those models can be directly linked to motor neuron models either phenomenologically \cite{Heckman1991, Fuglevand1993} or biophysically \cite{Cisi2008, Negro2011}

to further investigate the relationship between neural and mechanical behavior. 
The desired degree of detail and complexity within these models requires the coupling of different physical phenomena on different temporal and spatial scales, e.g., models describing the mechanical or electro-physiological state of the muscle tissue on the organ scale and the bio-chemical processes on the cellular scale  (cf. Section \ref{sec:model}).  

Being able to take into account all these different processes on different scales requires a flexible multi-scale, multi-physics computational framework and significant compute power. The availability of computational resources restricts the number of individual muscle fibers that can be considered within a skeletal muscle. The chemo-electromechanical models as implemented within the international open-source libraries OpenCMISS \cite{Bradley2011, Heidlauf2013,Mordhorst2015} allow general muscle geometries with about 1000 embedded ``muscle fibers''. Most skeletal muscles, however, have significantly more fibers. While simulations with 1000 fibers and less can potentially provide some insights into the neuromuscular system, some effects such as the motor unit recruitment over the full range of motor units and muscle fibers and their implication on the resulting EMG can not be estimated unless the full model with a realistic number of muscle fibers is simulated. This full model allows us to estimate the accuracy of ``reduced'' models by comparing them to the output of the detailed full ``benchmark'' model. Unless such comparisons are carried out, only predictions can be made how additional details such as more fibers or functional units (motor units) affect the overall outcome -- both in terms of muscle force generation and in terms of computed EMG signals.

While highly optimized and highly parallel software for biomechanical applications exists, e.g., for chemo-electromechanical heart models \cite{Gurev2015}, most multi-purpose computational frameworks for biomedical applications such as OpenCMISS are developed to provide flexibility. This is, e.g., achieved through standards like CellML \cite{Lloyd2004} and FieldML \cite{Christie2009}. The respective frameworks are utilised to enhance existing multi-physics models for a wide range of (bioengineering) applications. They are designed to be run by biomedical researchers on small-sized compute clusters. While they typically can be compiled on large-scale HPC compute clusters such as HazelHen at the HLRS in Stuttgart, they typically are not capable of exploiting the full potential of the machine. Moreover, simulation run time is typically considered less important than model complexity and output. Hence, typical simulations of biomedical applications are not necessarily optimized for numerical efficiency, minimal communication, the exploitation of novel algorithms or file output. Within this paper, we demonstrate how one can exploit analysis tools, suitable numerical techniques, and coupling strategies to obtain an efficient chemo-electromechanical skeletal muscle model that is suitable to be run on a large-scale HPC infrastructure and, thus, is capable of running with a sufficient resolution and number of muscle fibers to provide the required details. Once large-scale simulations of biomedical applications are solved with a high degree of detail, most specialized visualization tools such as OpenCMISS-Zinc, can no longer handle the amount of output data. Dedicated visualization tools for large-scale visualizations need to be developed. In this work, the MegaMol framework \cite{Grottel2015} has been adapted to visualize the different biophysical parameters and the resulting EMG.

\section{The multi-scale skeletal muscle model}\label{sec:model}
Before outlining the model in its full detail, we first provide a brief overview on some anatomical and physiological characteristics of skeletal muscles that are relevant to our model. From an anatomical point of view, skeletal muscles are a hierarchical system. Starting from its basic unit, the so-called sarcomere, several in-series and in-parallel arranged sarcomeres constitute a cylindrically shaped myofibril. Several in-parallel arranged myofibrils make up a skeletal muscle fiber and multiple muscle fibers form a fascicle. All fascicles constitute the entire muscle. These structures are connected with each other through an extracellular matrix (ECM). From a physiological point of view, several fibers are controlled by the same lower motor neuron through nervous axons. The entire unit consisting of the lower motor neuron, the axons and the respective fibers that are innervated by the axons, is referred to as motor unit, which is the smallest unit within a skeletal muscle that can voluntarily contract. The lower motor neuron sends rate-coded impulses called action potentials to all fibers belonging to the same motor unit. Moreover, motor units are activated in an orderly fashion, starting from the smallest ones, up to the biggest (recruitment size principle). 
After a motor neuron stimulates a muscle fiber at the neuromuscular junction, an action potential (AP) is triggered and propagates along the muscle fiber resulting in a local activity.
For more comprehensive insights into muscle physiology and anatomy, we refer to the book by \cite{MacIntosh2006}.

As the focus of this research is on solving biophysically detailed models of a neuromuscular system on HPC architectures, this section provides an overview on the multi-scale modeling framework of our chemo-electromechanical skeletal muscle model that is based on the work by \cite{Roehrle2012}, \cite{Heidlauf2013,Heidlauf2014}, and \cite{Heidlauf2016}. These models can account for the main mechanical and electro-physiological properties of skeletal muscle tissue, including a realistic activation process and resulting force generation. 
This is realized by linking multiple sub-models, describing different physical phenomena.  
To reduce the computational costs, the different sub-models that describe phenomena on different length and time scales are simulated using different discretizations, \iec spatial resolution and time-step size. Data are exchanged between the sub-models using homogenization and interpolation techniques. 

\subsection{The 3D continuum-mechanical muscle model}
The physiological working range of skeletal muscles includes large deformations. 
Therefore, we use a continuum mechanical modeling approach that is based on the theory of finite elasticity to simulate the macroscopic deformations and stresses in the muscle tissue.
In continuum mechanics, the placement function $\chi$ describes the motion of a material point, 
\iec it assigns every material point with position $\boldsymbol{X}$ in the reference (non-deformed) domain $\Omega_0 \subset \mathbb{R}^3$ at a time $t_0$ to a position $\boldsymbol{x}=\chi(\boldsymbol{X},t)$ in the actual (deformed) domain $\Omega_t \subset \mathbb{R}^3$ at time $t$.
The deformation of the body at a material point can be described by the deformation gradient tensor 
\[ \boldsymbol{F} := \frac{\partial \chi}{\partial \boldsymbol{X}}=\frac{\partial \boldsymbol{x}}{\partial \boldsymbol{X}}, \] 
which is defined as the partial derivative of the placement function $\chi$ with respect to the reference configuration.
The local displacement is defined by the vector $\boldsymbol{u}=\boldsymbol{x}-\boldsymbol{X}$.
  
The governing equation of the continuum mechanical model is the balance of linear momentum.  
Under the assumption of slow motions (\ie inertia forces vanish) and neglecting body forces, the balance of linear momentum in its local form can be written as 
\begin{equation}
  \operatorname{div} \boldsymbol{P} \ = \ \boldsymbol{0} \, \mbox{ in } \, \Omega_t \, \mbox{ for all } t, \label{e:3D} 
\end{equation}
where $\operatorname{div}(\cdot)$ denotes the divergence operator and $\boldsymbol{P}$ is the first Piola-Kirchhoff stress-tensor. 
To solve the balance of linear momentum, one needs to define a constitutive equation that specifies $\boldsymbol{P}$. It describes the overall mechanical behavior of the muscle that can be divided into a passive and an active component. The latter represents the muscle's ability to contract and produce forces.
In this work, we assume a superposition of the active and passive behavior, \iec an additive split of $\boldsymbol{P}$.

Passive skeletal muscle tissue is assumed to be hyperelastic and transversely isotropic. 
Consequently, the passive part to the first Piola-Kirchhoff stress tensor $\boldsymbol{P}_\mathrm{passive}(\boldsymbol{F},\boldsymbol{\mathcal{M}}$) depends on the deformation gradient tensor $\boldsymbol{F}$ 
and a structure tensor $\boldsymbol{\mathcal{M}}=\boldsymbol{a}_0 \otimes \boldsymbol{a}_0$, which is defined by the muscle fiber direction $\boldsymbol{a}_0$. 
The isotropic part of the passive stress-tensor assumes a Mooney-Rivlin material. It is enhanced by an additive anisotropic contribution accounting for the specific material properties in the muscle fiber direction $\boldsymbol{a}_0$.    

The active force is generated on a microscopic scale, \ie, within a half-sarcomere (the smallest functional unit of a muscle) consisting of thin actin and thick myosin filaments. 
Based on geometrical considerations of the half-sarcomere structure, it is known that the active muscle force depends on the actual half-sarcomere length $l_\mathrm{hs}$ (force-length relation) \cite{Gordon1966}.
When a half-sarcomere is activated by calcium as a second messenger, actin and myosin filaments can form cross-bridges and produce forces (cross-bridge cycling).

The active force state of the microscopic half-sarcomere is summarized in an activation parameter $\gamma$ that enters the macroscopic constitutive equation.
Furthermore, we assume that the active stress contribution acts only along the fiber direction $\boldsymbol{a}_0$.
When considering only isometric or slow contractions, the active stress tensor $\boldsymbol{P}_\mathrm{active}(\boldsymbol{F},\boldsymbol{\mathcal{M}},\gamma)$ can be defined as a function of the  lumped activation parameter $\gamma$, the deformation gradient tensor $\boldsymbol{F}$, and the structure tensor $\boldsymbol{\mathcal{M}}$. An additional force-length relationship needs to be included within $\boldsymbol{P}_\mathrm{active}$.

Finally, we assume skeletal muscle tissue to be incompressible, which implies the constraint $\operatorname{det} \boldsymbol{F} = 1$. The resulting first Piola-Kirchhoff stress tensor reads
\begin{equation}
  \boldsymbol{P} (\boldsymbol{F},\boldsymbol{\mathcal{M}},\gamma) \ = \ \boldsymbol{P}_\mathrm{passive} (\boldsymbol{F}, \boldsymbol{\mathcal{M}}) \, + \, \boldsymbol{P}_\mathrm{active} (\boldsymbol{F}, \boldsymbol{\mathcal{M}},\gamma) \, - \, p \boldsymbol{F}^{-T} \ ,
\label{equ:stress_tensor}
\end{equation}
where $p$ is the hydrostatic pressure, which enters the equation as a Lagrange multiplier enforcing the incompressibility constraint.
The material parameters of the continuum-mechanical skeletal muscles are fitted to experimental data \cite{Hawkins1994} and can be found in \cite{Heidlauf2014}.

\subsection{The 1D Model for Action Potential Propagation}
The electrical activity of skeletal muscles resulting from the local activity of all muscle fibers can be analyzed by measuring the extracellular potential.
The bidomain-model is a widely used continuum mechanical framework to simulate the electrical activity of living tissues \cite{Pullan2005}. It is based on the homogeneity assumption that intracellular and extracellular space homogeneously occupy the same domain.  
Intracellular and extracellular space are electrically coupled by an electrical current $I_\mathrm{m}$ flowing over the cell membrane,
\begin{equation*}
  \begin{array}{lll}
    - \operatorname{div} \ \boldsymbol{q}_\mathrm{i} = \operatorname{div} \ \boldsymbol{q}_\mathrm{e} = A_\mathrm{m} I_\mathrm{m},
  \end{array}
\end{equation*} where $\boldsymbol{q}_\mathrm{i}$ and $\boldsymbol{q}_\mathrm{e}$ denote the current density in the intracellular and extracellular space, respectively, and $A_\mathrm{m}$ is the fiber's surface to volume ratio. 
The muscle fiber membrane is nearly impermeable for ions and serves as a capacitor. However, ions can be transported over the membrane by ion channels and active ion pumps.
These properties can be mathematically described by a biophysically motivated modeling approach proposed by \cite{Hodgkin-Huxely-1952}, leading to the constitutive equation
\begin{equation}
I_\mathrm{m} \ = \ C_\mathrm{m} \frac{\partial V_\mathrm{m}}{\partial t} + I_\mathrm{ion}(\boldsymbol{y}, V_\mathrm{m}, I_\mathrm{stim}) \ ,
\label{eqn:monodomain_0D}
\end{equation} 
where $V_\mathrm{m}$ is the transmembrane potential, $C_\mathrm{m}$ is the capacity of the muscle fiber membrane (sarcolemma) and $I_\mathrm{ion}$ is the ionic current flowing over the ion-channels and -pumps, which depends on $V_\mathrm{m}$. Further state variables are summarized in $\boldsymbol{y}$, \egc the states of different ion channels. $I_\mathrm{stim}$ is an externally applied stimulation current, \egc due to a stimulus from the nervous system.
Assuming that the intracellular space and extracellular space show equivalent anisotropy, which is the case for 1D problems, the bidomain equations can be reduced to the monodomain equation. 
We use the one-dimensional monodomain equation in the domain $\Gamma_t\subset\mathbb{R}$:
\begin{equation}
\frac{\partial V_\mathrm{m}}{\partial t} = \frac{1}{A_\mathrm{m} C_\mathrm{m}} \left( \frac{\partial}{\partial x}\Big( \sigma_\mathrm{eff} \,\frac{\partial V_\mathrm{m}}{\partial x} \Big) - A_\mathrm{m} I_\mathrm{ion} \left(\boldsymbol{y}, V_\mathrm{m}, I_\mathrm{stim}\right)  \right) \, \mbox{ in } \, \Gamma_{t}. 
\label{eqn:monodomain}
\end{equation}         
Here, $x$ denotes the spatial coordinate along a one-dimensional line, i.e., the fiber, and $\sigma_\mathrm{eff}$ is the effective conductivity. 

\subsection{The 0D Sub-Cellular Muscle Model}\label{sec:cell_model}
The model proposed by \cite{Shorten2007} 
provides a basis to compute the lumped activation parameter $\gamma$ which is the link to the 3D continuum-mechanical muscle model. 
Its evolution model is steered by the ionic current $I_{\mathrm{ion}}$ of the 1D model.

It contains a detailed biophysical description of the sub-cellular excitation-contraction coupling pathway.
Specifically, it models the depolarization of the membrane potential in response to stimulation, release of calcium from the sarcoplasmic reticulum (SR) serving as a second messenger, and cross-bridge (XB) cycling. To do so, the Shorten model couples three sub-cellular models: A Hodgkin-Huxley-type model is utilized to simulate the electrical potentials and ion currents through the muscle-fiber membrane and the membrane of the T-tubule system. To simulate calcium dynamics, a model of the SR membrane ryanodine receptor (RyR) channels (\cite{Rios1993}) is coupled to the electrical potential across the T-tubule membrane and triggers the release of calcium from the SR. Additionally, the calcium-dynamics model describes diffusion of calcium in the muscle cell, active calcium transport over the SR membrane via the SERCA pump (sarco/endoplasmic reticulum $\mathrm{Ca}^{2+}$-ATPase), binding of calcium to buffer molecules (\eg parvalbumin or ATP), and binding of calcium to troponin enabling the formation of cross-bridges. 
The active force generation is simulated by solving a simplified Huxley-type model \cite{Razumova1999}, which is the basis for calculating the activation parameter $\gamma$. 

The entire incorporated sub-cellular processes are modeled with a set of coupled ordinary differential equations (ODEs)
\begin{equation}
	\frac{\partial \boldsymbol{y}}{\partial t} \ =  G_{\boldsymbol{y}}
	\left( \boldsymbol{y}, V_\mathrm{m} ,  I_\mathrm{stim} \right)\, ,
	\label{eqn:0D_state}
\end{equation}
where $G_{\boldsymbol{y}}$ summarizes the right-hand-side of all the ODEs associated with the state variables $\boldsymbol{y}$ which are, in the case of the Shorten et al.~model, more than 50. 
The final activation parameter $\gamma$ is computed from entries of the state variable vector $\boldsymbol{y}$ and the length and contraction velocity of the half-sarcomere, $l_\mathrm{hs}$ and $\dot{l}_\mathrm{hs}$.
Assuming isometric or very slow contractions, the contraction velocity can be neglected. Hence, following \cite{Razumova1999} and \cite{Heidlauf2014}, the activation parameter is calculated as
\begin{equation}
 \gamma\left(\boldsymbol{y},l_\mathrm{hs}\right)  = f_\mathrm{f\text{-}l} \left(l_\mathrm{hs}\right) \, \dfrac{A_2- A_2^\text{min}}{A_2^\text{max}- A_2^\text{min}} \ .
 \label{equ:gamma}
\end{equation}
Here, the function $f_\mathrm{f\text{-}l} \left( l_\mathrm{hs}\right)$ is the force-length relation for a cat skeletal muscle by \cite{Rassier1999}, 
$A_2 \in \boldsymbol{y}$ is the concentration of post power-stroke XBs, $A_2^\text{max}$ is the concentration of post power-stroke cross-bridges for a tetanic contraction (100~Hz stimulation after 500~ms stimulation) and $A_2^\text{min}$ is an offset parameter denoting the concentration of post power-stroke cross-bridges in the resting state.

\subsection{Test Scenario, Boundary and Initial Conditions}\label{sec:test_scenario}

The overall aim of this work is to analyse the implemented algorithms on (large-scale) clusters and to identify bottlenecks that prevent us from achieving scalable and efficient implementations. For this purpose, we define a test scenario that will be used throughout this work to test the model dynamics and the implemented algorithms. 
To summarize the previous paragraphs, the chemo-electromechanical behavior of a skeletal muscle is described by the following coupled equations:
\begin{subequations}
\begin{align}
&\boldsymbol{0} \ & = & \  \operatorname{div} \boldsymbol{P}\big(\boldsymbol{F},\boldsymbol{\mathcal{M}},\gamma(\boldsymbol{y}, 
\l_\mathrm{hs}) \big) & \text{ \hspace{-2cm} in } \, \Omega_t \, \mbox{ for all } \; t , \label{e:3D_momentum_balance} \\
&\frac{\partial V_\mathrm{m}}{\partial t} \ & = & \ \frac{1}{A_\mathrm{m} C_\mathrm{m}} \left( \frac{\partial}{\partial x}\Big( \sigma_\mathrm{eff} \,\frac{\partial V_\mathrm{m}}{\partial x} \Big) - A_\mathrm{m} I_\mathrm{ion}\big(\boldsymbol{y}, V_\mathrm{m}, I_\mathrm{stim} \big) \right) & \text{ \hspace{-2cm} on all fibers } \, \Gamma_{t} \,, \label{e:1D_fiber} \\
&\frac{\partial \boldsymbol{y}}{\partial t} \ & = & \ G_{\boldsymbol{y}} \left(\boldsymbol{y}, V_\mathrm{m} , I_\mathrm{stim} \right) & \mbox{ \hspace{-2cm} at all half-sacromere positions.} \label{e:0D_ystate2} 
\end{align}\label{eq_summary}
\end{subequations}

For the test scenario, we use a generic cubic muscle geometry ($1\, \text{cm} \times 1 \, \text{cm} \times 1 \, \text{cm}$). The muscle fibers are aligned parallel to one cube-edge (\ie the $e_x$-direction). We consider a $10.0\,\text{ms}$-long isometric single-twitch experiment by stimulating all fibers at their mid-points for $t\in[0, 0.1 \,\text{ms}]$ with $I_\mathrm{stim}=1200\, \upmu \textrm{A/}\textrm{cm}^2$.
To constrain the generic muscle sample, Dirichlet boundary conditions (zero displacement) are used resulting in the fixation of the following faces of the muscle cube: the left and the right faces (faces normal to the $e_x$-direction), the front face (face normal to the $e_y$-direction) and the bottom face (face normal to the $e_z$-direction). 
Further, no current flows over the boundary of the computational muscle fibers, \iec zero Neumann boundary condition are assumed at both muscle fiber ends.

For the material parameters for the continuum-mechanical model, the effective conductivity $\sigma_\text{eff}$, the surface-to-volume ratio $A_\mathrm{m}$, and the membrane capacity $C_\mathrm{m}$, we use exactly the same values as reported in \cite{Heidlauf2014}.  
The initial values for the state vector $\boldsymbol{y}$ at $t_0$ of the sub-cellular model are taken from the slow-twitch scenario of \cite{Shorten2007}. 

\subsection{Status Quo and Current Implementation}
\label{sec:StatusQuo}

\textit{Spatial Discretization.}
The sub-models of the multi-scale skeletal muscle model show significantly different characteristic time and length scales.
In order to solve the model efficiently, different discretization techniques and resolutions are required for the sub-models. The current implementation of the OpenCMISS solves the continuum-mechanical skeletal muscle model using a finite element method with triquadratic/trilinear Lagrange basis functions, \iec Taylor-Hood elements.  
The one-dimensional muscle fibers are represented by much finer one-dimensional finite element meshes with linear Lagrange basis functions. 
The respective one-dimensional elements are embedded into the three-dimensional finite element mesh.

\begin{figure}[h!]
  \includegraphics[width=0.98\textwidth]{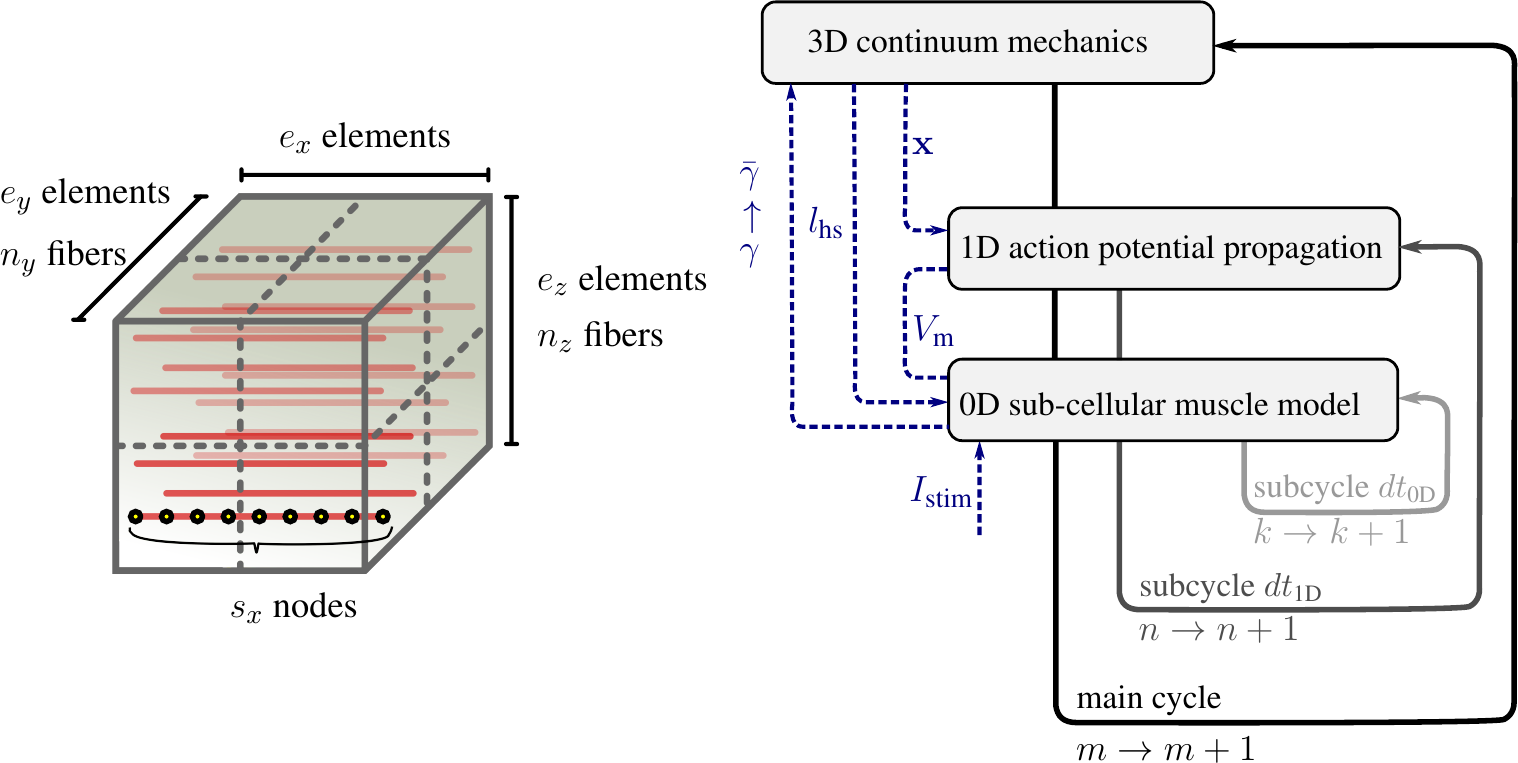}
	\caption{Left: Schematic view of a 3D muscle domain that contains a given number of $n_x \times n_y$ muscle fibers per 3D partition,
$e_x \times e_y \times e_z$ finite elements for the 3D model \eqref{e:3D_momentum_balance}, and $s_x$ nodes per fiber for
\eqref{e:1D_fiber} and \eqref{e:0D_ystate2}. Right: Schematic view of the multi-scale time stepping scheme based on a Godunov splitting of the monodomain equation.} 
	\label{fig:muscle_cube}
\end{figure}

To simulate the coupled multi-physics problem, data have to be transfered between the different spatial discretization approaches.
The microscopic sarcomere forces $\gamma$ provided by the monodomain model are homogenized and projected to the macroscopic Gauss points
of the three-dimensional continuum-mechanical model ($\gamma \rightarrow \bar{\gamma}$). 
Vice versa, the node positions of the one-dimensional computational muscle fibers are updated from the actual displacements $\boldsymbol{u}$ of the three-dimensional continuum-mechanical model by interpolating the node positions via the basis-functions of the three-dimensional model. Based on this step, the microscopic half-sarcomere lengths $l_\mathrm{hs}(\boldsymbol{x})$ are evaluated. Fig.~\ref{fig:muscle_cube} (left) schematically shows the embedding of $n_y \times n_z$ discretised 1D fibers within the 3D muscle domain $\Omega_0$ discretised with $e_x \times e_y \times e_z$ finite elements.

\textit{Time discretization and coupling in time.} To compute an approximate solution for Eqns \eqref{eq_summary}, the different characteristic time scales of the 3D, 1D and 0D problems can be exploited. The action potential propagates faster than the muscle deformation, 
the sub-cellular chemo-mechanical processes evolve considerably faster than the diffusive action potential propagation. 
To achieve common global time steps, which is desirable from a computational point of view, we choose $dt_\mathrm{3D}/\mathrm{N}= dt_\mathrm{1D}= \mathrm{K} \cdot dt_\mathrm{0D}$, with $\mathrm{N},\mathrm{K} \in \mathbb{N}$ and define $t_{m,n,k} :=  m \cdot dt_\mathrm{3D} + n \cdot dt_\mathrm{1D} + k \cdot dt_\mathrm{0D}$. State values associated with time $t_{m,n,k}$ are denoted with the superscript $(\cdot)^{m,n,k}$. Employing different time steps requires a time splitting scheme. The status-quo implementation uses a first-order accurate Godunov splitting scheme, for which one time-step of the three-dimensional equation including all sub-steps for the one-dimensional monodomain equation reads:

\begin{enumerate}
\item For $n=0,\ldots,N-1$ do
      \begin{enumerate}
      \item For $k=0,\ldots,K-1$ perform explicit Euler steps for Eqn.~\eqref{e:0D_ystate2} and the 0D portion of Eqn.~\eqref{e:1D_fiber}. 
      \item Set $V_\mathrm{m}^{m,n,0}:=V_\mathrm{m}^{m,n,K}$ and $\boldsymbol{y}^{m,n+1,0} := \boldsymbol{y}^{m,n,K}$.     
      \item Perform one implicit Euler step for the 1D portion of Eqn.~\eqref{e:1D_fiber} to compute $V_\mathrm{m}^{m,n+1,0}$.
      \end{enumerate}
\item Set $V_\mathrm{m}^{m+1,0,0}:=V_\mathrm{m}^{m,N,0}$ and $\boldsymbol{y}^{m+1,0,0} := \boldsymbol{y}^{m,N,0}$. 
\item Calculate $\gamma (\boldsymbol{y}^{m+1,0,0}, l_\mathrm{hs}^{m,0,0})$  and compute the homogenized values $\bar{\gamma}$ at the Gauss points of the 3D finite element mesh.
\item Solve Eqn.~\eqref{e:3D_momentum_balance}.
\item Interpolate the actual configuration $\boldsymbol{x}^{m+1,0,0}$ to the fibers' nodes for computing the local half-sacromere length $l_\mathrm{hs}^{m+1,0,0}$.
\end{enumerate}

Fig.~\ref{fig:muscle_cube} (right) schematically depicts the algorithm. We assume quasi-static conditions for the continuum-mechanical problem. This is equivalent to assuming constant muscle deformations $\boldsymbol{u}$ throughout $dt_\mathrm{3D}$. 

\textit{Solvers.} The coupled time stepping algorithm described above contains two large systems of equations that need to be solved.
 The first one results from the 3D elasticity problem \eqref{e:3D_momentum_balance} and the second one stems from an implicit time integration of the linear 1D diffusion problem of the fiber \eqref{e:1D_fiber}. In the existing implementation, the GMRES solver of \cite{GMRES} from the PETSc\footnote{\tt http://www.mcs.anl.gov/petsc/index.html} library is used. 

\textit{Domain Partitioning.} 
The 3D computational domain is decomposed into disjoint cuboid shaped partitions by cutting the whole domain with axis-aligned planar cuts at element boundaries. The partitioning spans all 
model components, such that the quantities in the 3D, 1D and 0D model corresponding to the same spatial location
are on the same process. This is to avoid unnecessary inter-process volume-communication between the sub-models.
The implementation of this partitioning concept is motivated by a skeletal muscle anatomy and physiology: all skeletal muscle fibers are, from an electrical point of view, independent from each other. 
Thus, the obvious way to partition the workload is to distribute whole fibers among the processes. 
For instance, \cite{Heidlauf2013} hard-coded a partitioning with 4 processes.

\section{Towards HPC: Scalability Evaluation}\label{sec:hpc}


Before being able to simulate such complex models as the one introduced in Section \ref{sec:model} on HPC systems,  
it is essential to first analyse bottlenecks of existing implementations.
In this paper, we focus on investigating numerical and algorithmic characteristics of the implementation as described in Sec.~\ref{sec:StatusQuo}. The results provide the basis for new algorithms and improved implementations to enhance scalability. As the parallel status quo code used 4 cores, we only analyse numerical complexity, \iec scalability in terms of the size of the problem, and redesign the parallelisation.

As discretization, we choose 8 Taylor-Hood finite elements
and include 36 muscle fibers ($n_x=n_y=6$), which go through all of the Gauss points that exhibit the same $x$- and $y$-coordinates. 
The time steps are $dt_\mathrm{3D}=1$, $dt_\mathrm{1D}=5\cdot10^{-4}\,\text{ms}$ and $dt_\mathrm{0D}=10^{-4}\,\text{ms}$, \iec $N=2000$ and $K=5$. 
The system resulting from the discretization of the 1D fibers is solved using a restarted GMRES solver with restart after 30 iterations and relative residuum tolerance $10^{-5}$. To solve the 3D problem, Newton's method from the PETSc library is used with relative and absolute tolerance $10^{-8}$ and a backtracking line search approach with maximum number of 40 iterations.

\begin{figure}[h!]
  \begin{center}
    \includegraphics[width=0.9\textwidth]{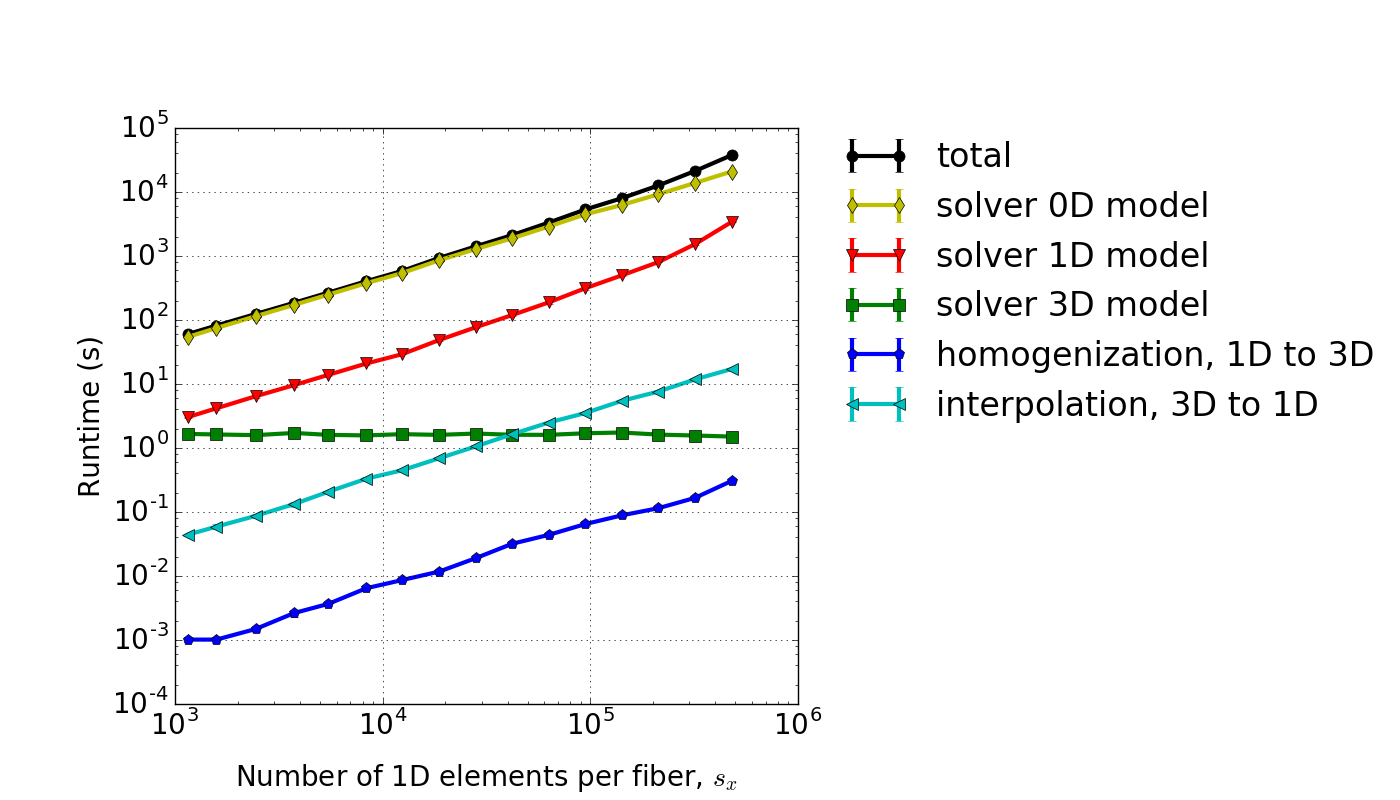}
  \end{center}
  \caption{Runtime for a simulated time interval $t\in[0,3\,\text{ms}]$ with varying number of elements per fiber.}
  \label{fig:serial_scaling}
\end{figure}

To assess problem size scalability, we vary the number of nodes along each muscle fiber and measure runtimes for a single step of the 3D model. 
Results are depicted in Fig.~\ref{fig:serial_scaling}. 

In a second experiment, 
we compare the previously used GMRES solver with a conjugate gradient (CG) solver and the direct MUMPS LU solver \cite{MUMPS01,MUMPS02} as implemented within PETSc \cite{petsc-efficient}.

\begin{figure}[h!]
  \centering
  \includegraphics[width=0.9\textwidth]{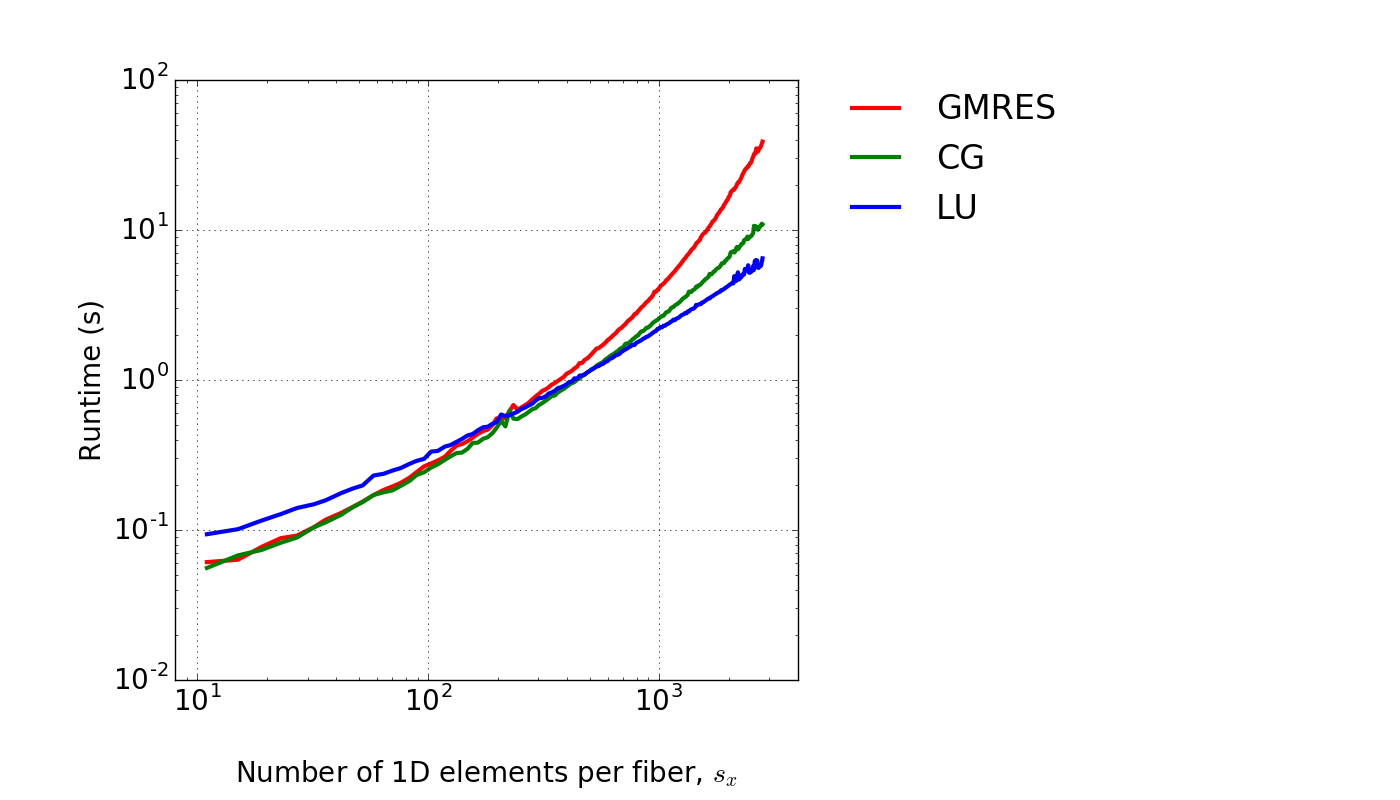}
  \caption{Comparison of runtime for different linear solvers and considering a single fiber and for a simulated time interval $t\in[0,3\,\text{ms}]$ with varying number of elements per fiber.}
  \label{fig:1d_solver}
\end{figure}

Overall, the results for both numerical investigations reveal the following insights:
\begin{enumerate}
  \item The   low-order explicit ODE solver (Euler) requires very small time steps for the desired accuracy. Hence, the majority of the runtime is spent for solving the 0D problem.
  \item The portion of runtime for solving the 3D problem is negligible and constant. This is due to the low number of 3D finite elements for the mechanics problem (eight elements). Realistic models would require a finer resolution of the 3D problem.
  \item The runtime for the other computational components increases approximately linearly with the number of fiber elements. This indicates a good scaling behavior with respect to problem size.

  \item The computation of the macroscopic variable $l_\mathrm{hs}$ from fiber nodes and the homogenized activation parameter $\bar{\gamma}$ has almost no impact on the computational time. Computing $l_\mathrm{hs}$ is more time consuming as it involves simultaneously traversing the fiber and the 3D meshes, whereas computing $\bar{\gamma}$ requires a single averaging operation for each Gauss point of the 3D elements.

  \item The GMRES solver is a robust choice for general sparse linear systems of equations, but does not exploit the symmetry, positive definiteness and tri-diagonal structure of a 1D diffusion system and hence exhibits substantial overhead. The LU solver exhibits the lowest computational time for large problems. It shows linear complexity in the number of 1D nodes\footnote{Note that the 1D diffusion is an exception. In all higher dimensions, direct solvers are known to fail for large problems due to their high complexity.}. 
\end{enumerate}

These findings are used in the following sections to improve the numerical efficiency from an algorithmic and implementational point of view. 

\section{Towards HPC: Improvements For efficiency}

Although linear scalability with respect to the number of nodes per fiber is observed for all 0D and 1D components (cf. Sec.~\ref{sec:hpc}), the runtime is not optimal. This is mainly due to the first-order accurate $(\mathcal{O}(dt))$ time stepping scheme. It enforces very small 
time steps (for $dt_\mathrm{1D}$ and $dt_\mathrm{0D}$) in order to achieve the required accuracy. We propose a combination of second-order splitting and second-order time stepping schemes within the 0D and 1D sub-problems to tackle this issue. A further obstacle on the way to large-scale simulations with
a reasonable number of fibers is the limited parallelism (currently using $4$ cores). We design a parallelization scheme for arbitrary core numbers along 
with a new partitioning scheme minimizing comunication cost. Finally, we replace the customized OpenCMISS file output and visualization by a new library 
that is able to handle very large-scale data.

\subsection{Algorithmic Optimization-- Second-Order Time Stepping}
\label{sec:numerics_opt}
To reduce computational cost and increase accuracy, we propose to replace the Godunov splitting with a second-order Strang splitting. We replace the explicit Euler for Eqn.~\eqref{e:0D_ystate2} and the 0D portion of Eqn.~\eqref{e:1D_fiber} with the method of Heun and employ an implicit Crank-Nicolson method for the diffusion part of Eqn.~\eqref{e:1D_fiber}. Second-order time-stepping schemes $(\mathcal{O}(dt^2))$ allow to reduce the discretization error much faster with decreasing time step size $dt$. Thus, the required accuracy might be achieved by employing larger time steps. 
In contrast to the simpler Godunov splitting, the Strang splitting uses three sub-steps per time step: a first step with length
$dt_\mathrm{1D}/2$ for the 0D part, a second step with length $dt_\mathrm{1D}$ for the diffusion, and a third step with length
$dt_\mathrm{1D}/2$ again for the 0D part. The modified algorithm at a time $t_{m,0,0}$ reads:

\begin{enumerate}
\item For $n=0,\ldots,N-1$ do
      \begin{enumerate}
      \item For $k=0,\ldots,K/2-1$ perform explicit Heun steps for Eqn.~\eqref{e:0D_ystate2} and the 0D portion of Eqn.~\eqref{e:1D_fiber}.      
      \item Set $V_\mathrm{m}^{m,n,0}:=V_\mathrm{m}^{m,n,K/2}$.
      \item Perform one implicit Crank-Nicolson step for the 1D portion of Eqn.~\eqref{e:1D_fiber}.
      \item Set $V_\mathrm{m}^{m,n,K/2}:=V_\mathrm{m}^{m,n+1,0}$.
      \item For $k=K/2,\ldots,K-1$ perform explicit Heun steps for Eqn.~\eqref{e:0D_ystate2} and the 0D portion of Eqn.~\eqref{e:1D_fiber}. 
      \item Set $V_\mathrm{m}^{m,n+1,0}:=V_\mathrm{m}^{m,n,K}$ and $\boldsymbol{y}^{m,n+1,0} := \boldsymbol{y}^{m,n,K}$. 
      \end{enumerate}
\item Set $V_\mathrm{m}^{m+1,0,0}:=V_\mathrm{m}^{m,N,0}$ and $\boldsymbol{y}^{m+1,0,0} := \boldsymbol{y}^{m,N,0}$. 
\item Calculate $\gamma (\boldsymbol{y}^{m+1,0,0}, l_\mathrm{hs}^{m,0,0})$  and compute the homogenised values $\bar{\gamma}$ at the Gauss points of the 3D finite element mesh.
\item Solve Eqn.~\eqref{e:3D_momentum_balance}.
\item Interpolate the displacements $\boldsymbol{u}^{m+1,0,0}$ to the fibers' nodes for computing the local half-sacromere length $l_\mathrm{hs}^{m+1,0,0}$.
\end{enumerate}

The explicit Heun step in a. and d. reads:
\begin{subequations}
			\begin{eqnarray}
			\left( \! \begin{array}{c} \boldsymbol{y} \\  V_\mathrm{m} \end{array}  \! \right)^{\mathrm{pre}} \! \!
			\! \! \! & = & \! \!\left(  \! \begin{array}{c} \boldsymbol{y} \\  V_\mathrm{m} \end{array} \! \right)^{m,n,k} \!
 			\! + dt_{\mathrm{0D}} \left( \! \begin{array}{l}
 			G_{\boldsymbol{y}} \left( \boldsymbol{y}^{m,n,k}, V_\mathrm{m}^{m,n,k}, I_\mathrm{stim} \right) \\
			- \dfrac{1}{C_\mathrm{m}} I_\mathrm{ion} \left( \boldsymbol{y}^{m,n,k}, V_\mathrm{m}^{m,n,k},  I_\text{stim}  \right) \end{array}  \! \right),  \label{eqn:Heun_1} \\
			\left( \! \begin{array}{c}  \boldsymbol{y}  \\  V_\mathrm{m}\end{array} \! \right)^{m,n,k+1} \! \!
			\! \! \! & = & \! \!\left(  \! \begin{array}{c}  \boldsymbol{y} \\ V_\mathrm{m} \end{array} \! \right)^{m,n,k} \label{eqn:Heun_2}  \\
			&& + \frac{dt_{\mathrm{0D}}}{2} \left( \! \begin{array}{c} 
			G_{\boldsymbol{y}} \left(\boldsymbol{y}^{m,n,k}, V_\mathrm{m}^{m,n,k}, I_\mathrm{stim} \right)
			+ G_{\boldsymbol{y}} \left( \boldsymbol{y}^\mathrm{pre}, V_\mathrm{m}^{\mathrm{pre}} , I_\mathrm{stim}\right) \\
			- \dfrac{1}{C_\mathrm{m}} \left( 
			I_\mathrm{ion} \left( \boldsymbol{y}^{m,n,k}, V_\mathrm{m}^{m,n,k}, I_\text{stim}\right) +
			I_\mathrm{ion} \left( \boldsymbol{y}^\mathrm{pre}, V_\mathrm{m}^{\mathrm{pre}}, I_\text{stim} \right) \right) \end{array} \! \right). \notag
			\end{eqnarray} 
\end{subequations}

In b., we solve the system resulting from the Crank-Nicolson time discretization of the diffusion part in Eqn.~\eqref{e:1D_fiber}.:
			\begin{equation}
			V_\mathrm{m}^{m,n+1,0} = V_\mathrm{m}^{m,n,0} + \frac{dt_\mathrm{1D}}{2 \, A_{\mathrm{m}} C_{\mathrm{m}}}
			\left( 
			\frac{\partial}{\partial x}\left( \sigma_\mathrm{eff} \,\frac{\partial V_\mathrm{m}^{m,n,0}}{\partial x} \right) + 
			\frac{\partial}{\partial x}\left( \sigma_\mathrm{eff} \,\frac{\partial V_\mathrm{m}^{m,n+1,0}}{\partial x} \right) \right),
			\label{eqn:CN}
			\end{equation}

In the following, we present numerical experiments demonstrating the increase in 
efficiency achieved as a result of this new time discretization.

\subsubsection{Time Discretization for the Sub-Cellular Model (0D ODEs)}
\begin{figure}[h!]
\centering
  \begin{minipage}[t]{0.47\linewidth}
    \includegraphics[width=\textwidth, trim={0.5cm 0cm 4cm 2cm},clip]{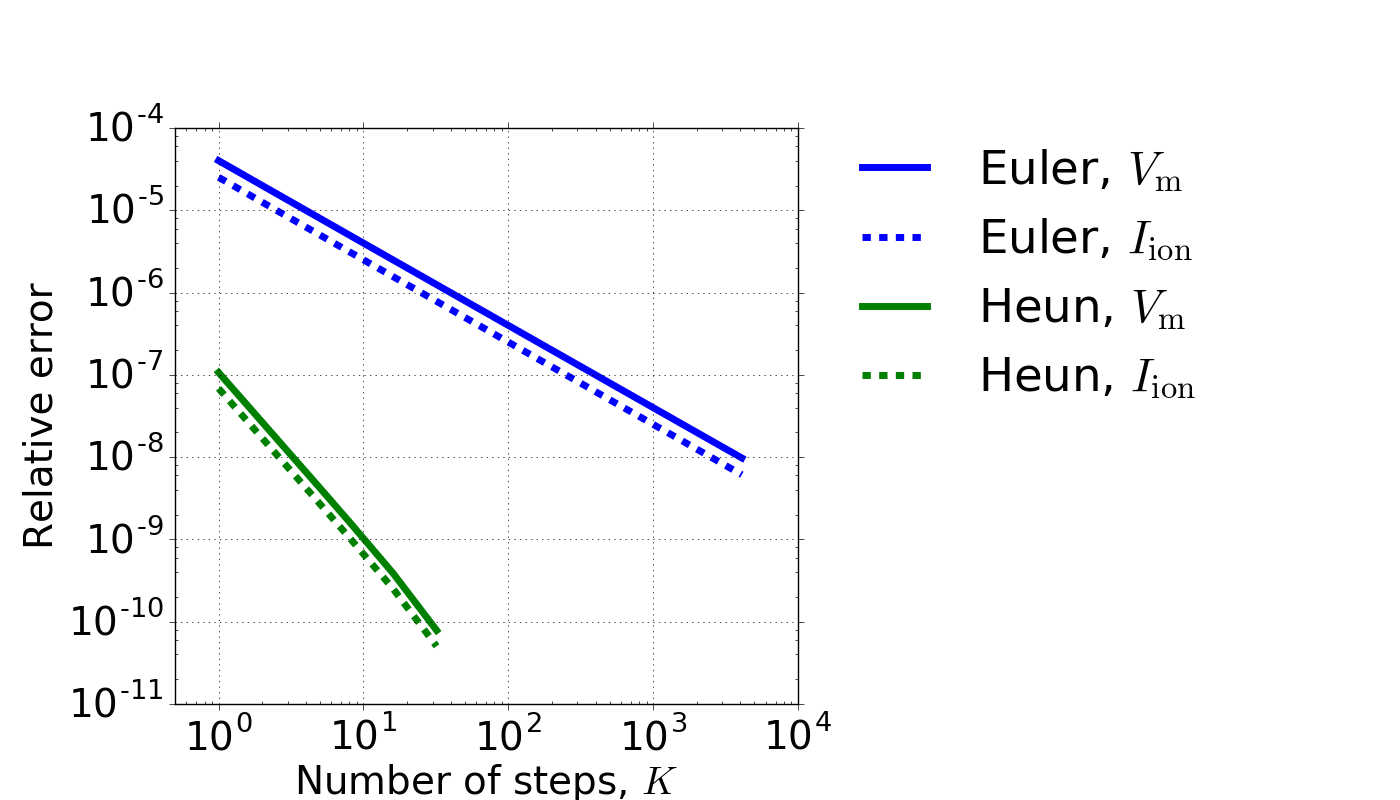}
    \caption{Relative error dependency on the number $K$ of $0D$ time steps in $[0,dt_\mathrm{1D}]$. The error of Euler's and Heun's method shows the expected $\mathcal{O}(K^{-1})$ and $\mathcal{O}(K^{-2})$ behavior.
    }
    \label{fig:0D_convB}
  \end{minipage}
  \hfill
  \begin{minipage}[t]{0.47\linewidth}
    \includegraphics[width=\textwidth, trim={0.5cm 0cm 4cm 2cm},clip]{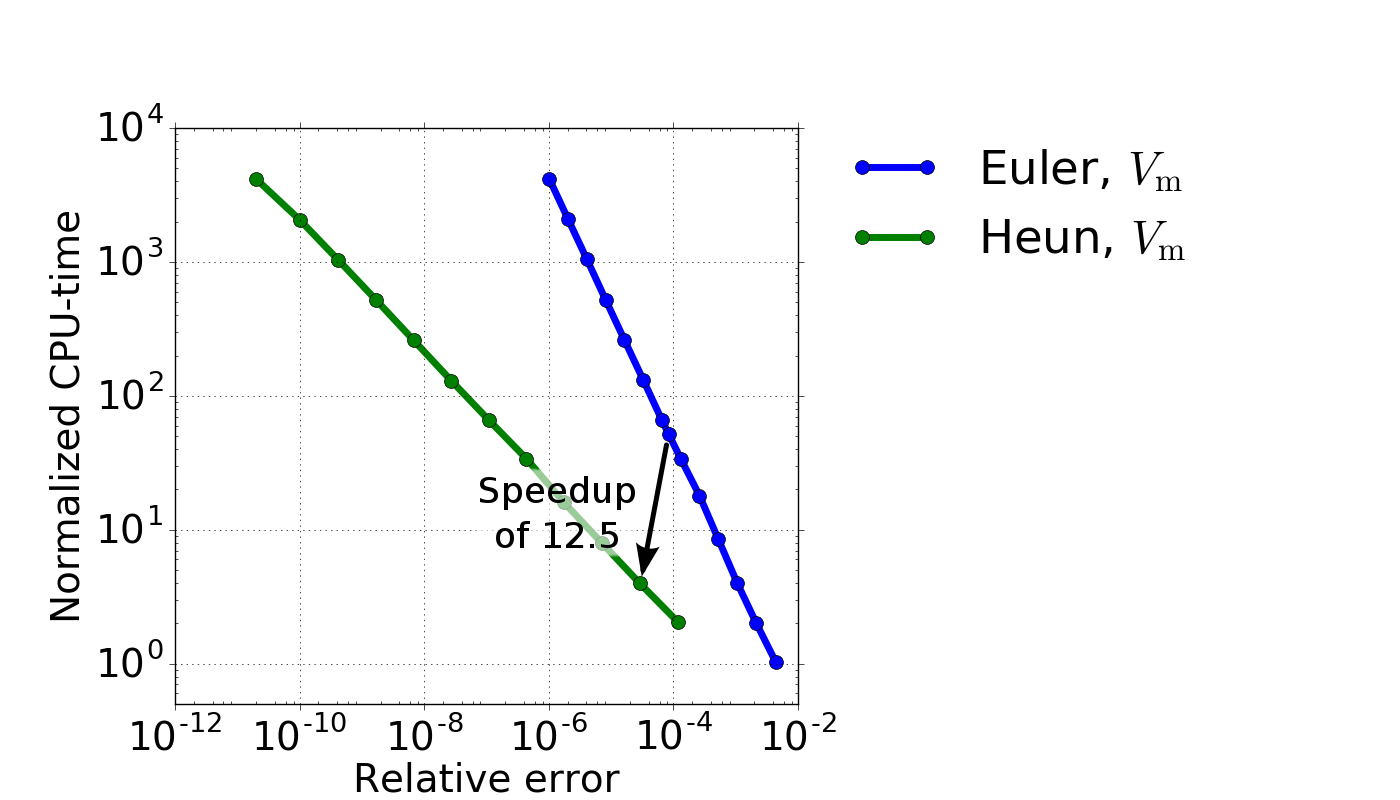}
    \caption{Dependency of the runtime on the required accuracy for explicit Euler and Heun. We varied the time step $dt_\mathrm{0D}$    
    between $5\cdot 2^0$ and $5\cdot 2^{12} \,\upmu\textrm{s}$ for
    Euler and between $5\cdot 2^0$ and $5\cdot 2^{11} \,\upmu\textrm{s}$ for
    Heun.}
    \label{fig:Efficiency50}
   \end{minipage}
\end{figure}

In a first step, we verify the convergence order of Heun's method numerically. Therefore, we restrict ourselves to step 1.a of the Godunov algorithm, but use Heun's method for Eqn.~\eqref{e:0D_ystate2} and the 0D portion of Eqn.~\eqref{e:1D_fiber}.
We use the same test setup as presented in Sects.~\ref{sec:test_scenario} and \ref{sec:StatusQuo}.
 To compare the accuracy of Heun's method with the explicit Euler method, we compare the values of $V_\mathrm{m}$ and $I_\mathrm{ion}$ at a stimulated material point on a muscle fiber while varying the time step size $dt_\mathrm{0D}$. To compare the methods in terms of efficiency, we measure the related compute times.
We restrict ourselves to the time interval 
$[0,dt_\mathrm{1D}]$, with $dt_\mathrm{1D}=0.5 \,\upmu \textrm{s}$. 
 As a reference solution, we use the solution calculated with Heun's method with a very high
resolution ($K=4096$). 
The tests are performed sequentially on an Intel{\tiny \textsuperscript{\textregistered}} Core{\tiny \textsuperscript{TM}} i5-4590 CPU, using the OpenCMISS implementation. Fig.~\ref{fig:0D_convB} depicts the relative error depending on the number $K$
of $0D$ time steps executed in the interval $[0,dt_\mathrm{1D}]$ while Fig.~\ref{fig:Efficiency50} compares the necessary CPU-time to reach a certain accuracy for the different solvers.

Fig.~\ref{fig:0D_convB} shows the expected first-order convergence for the explicit Euler method and second-order convergence for Heun's method. 
From an application point of view, however, efficient computation (``Which accuracy can be achieved in which runtime?'') is more important than the order of convergence. 
Since each time step of the Heun's method is twice as expensive as an explicit Euler step, we need to also take into account the runtime in our assessment 
to show the potential to decrease the runtime for Heun's method given a required accuracy.
All times are normalized by the CPU-time of the Euler method with $K=1$. The results presented in Fig.~\ref{fig:Efficiency50} show that two Heun steps with
$dt_\mathrm{0D}=2.5\,\upmu\textrm{s}$ replace $50$ forward Euler steps yielding a theoretical speedup of $12.5$ for the 0D-solver. At the same time, the error decreases by a factor of approximately $3$.

\subsubsection{Time Discretization for Muscle Fibers (0D coupled to 1D)}
In the second step, we verify the convergence order of the Strang splitting scheme. We consider again the same test setup as above except that we use the larger time interval $[0,0.1\,\textrm{ms}]$.  
We varying the number $\mathrm{N}$ of 1D time steps. Based on the previous results for the isolated 0D problem, we choose $\mathrm{K}=2$ for the Strang-splitting scheme and $\mathrm{K}=5$ for the Godunov-splitting scheme. This ensures a comparable relative error for the 0D sub-problem while saving computational time. The reference solution is computed using the Strang-splitting scheme with $dt_\mathrm{1D}=0.25 \, \upmu \textrm{s}$, yielding $V_\mathrm{m}(0.1\,\textrm{ms}) \approx -23.5219 \,\textrm{mV}$. 

\begin{figure}[h!]
\centering
  \begin{minipage}[t]{0.47\linewidth}
    \includegraphics[width=\textwidth, trim={0.5cm 0cm 4cm 2cm},clip]{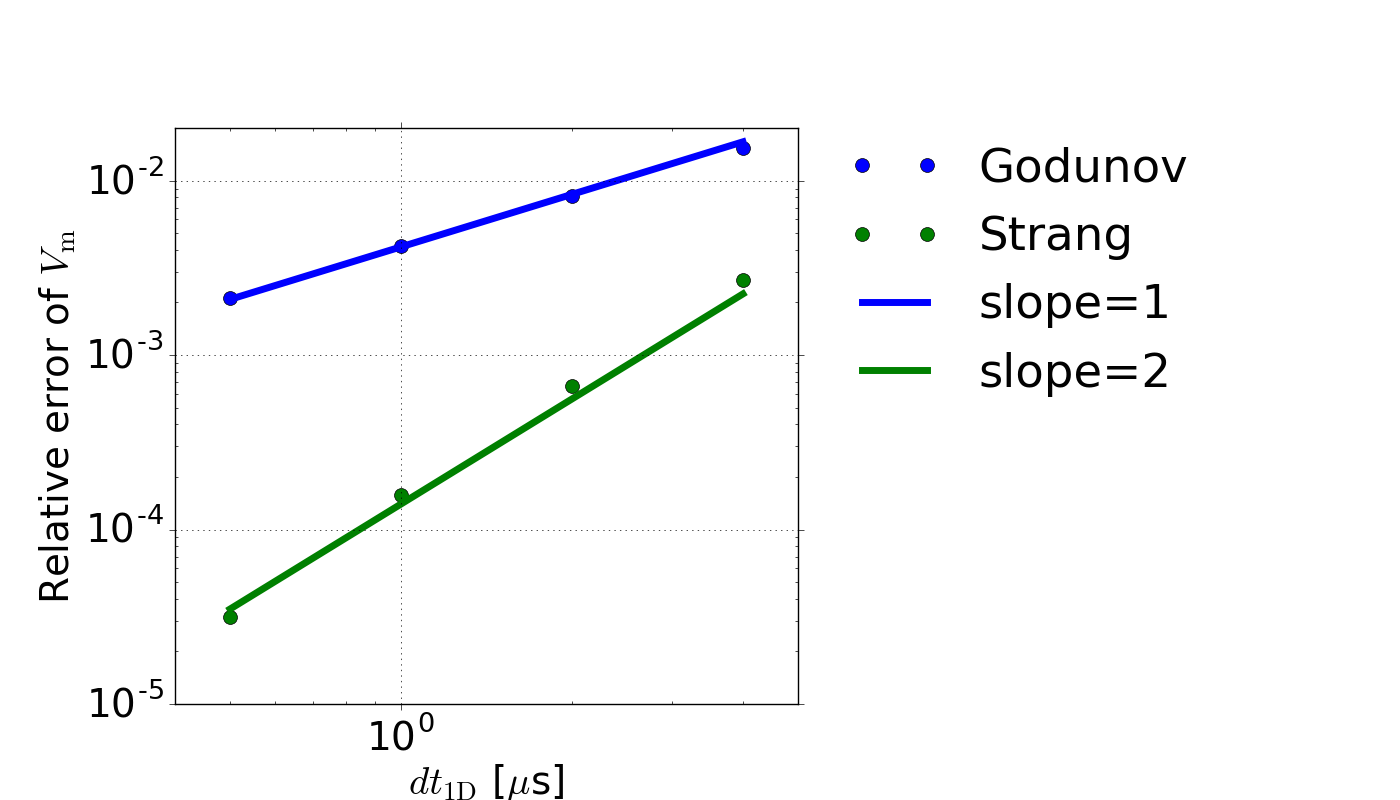}
    \caption{Relative error dependency on the 1D time step size $dt_\mathrm{1D}$. The error of the Godunov- and Strang-splitting scheme shows the expected $\mathcal{O}(dt_\mathrm{1D})$ and $\mathcal{O}(dt_\mathrm{1D}^2)$ behavior.}
    \label{fig:SplittingError}
  \end{minipage}
  \hfill
  \begin{minipage}[t]{0.47\linewidth}
    \includegraphics[width=\textwidth, trim={0.5cm 0cm 4cm 2cm},clip]{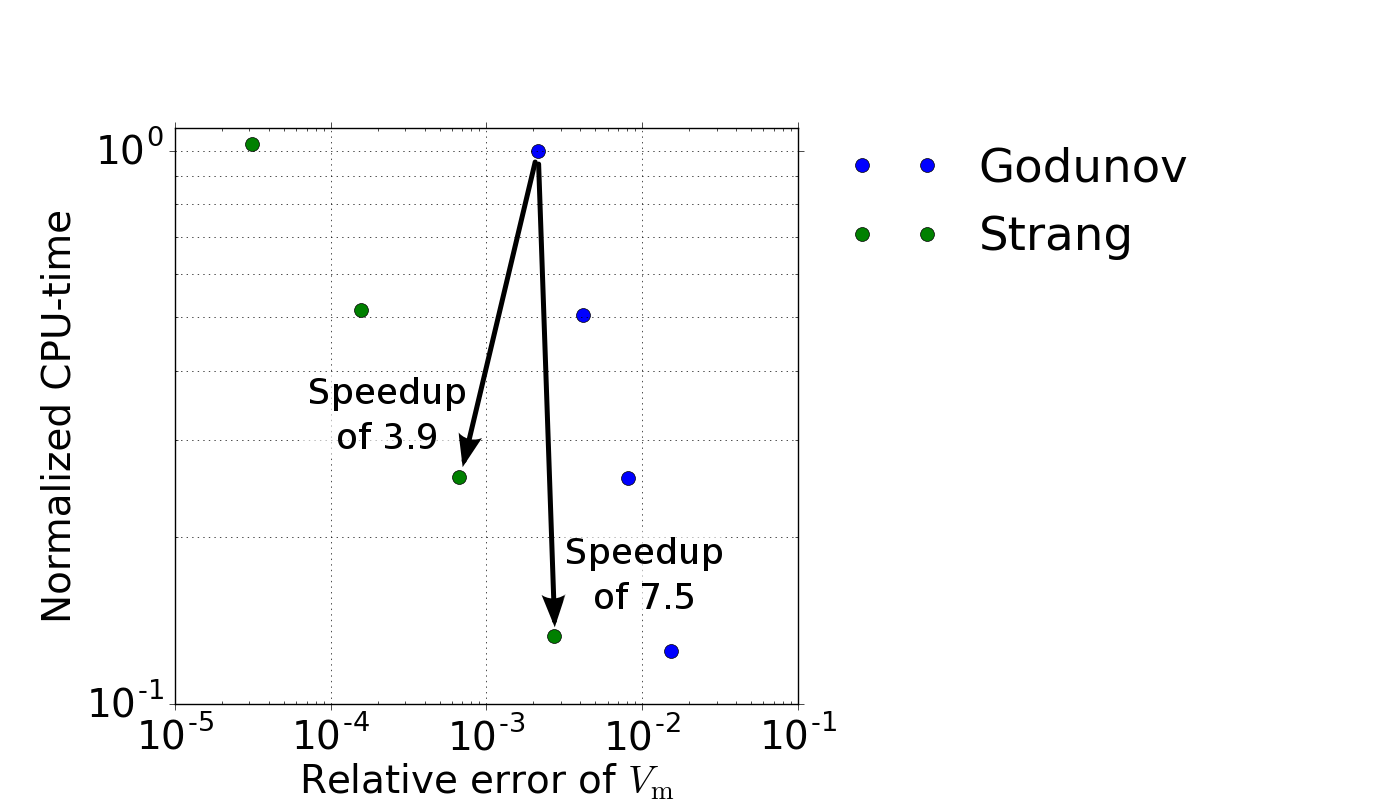}
    \caption{Efficiency of different splitting schemes. Each scheme is performed for $dt_\mathrm{1D}=0.5,1,2 \textrm{ and } 4\,\textrm{ms}$.}
    \label{fig:SplittingEfficiency}
   \end{minipage}
\end{figure}

Fig.~\ref{fig:SplittingError} shows the relative errors of $V_\mathrm{m}\left(0.1\,\textrm{ms}\right)$ at a stimulated sub-cell for the Godunov- and Strang-splitting schemes. Comparable relative errors as for the Godunov scheme with $dt_\mathrm{1D}=0.5\,\upmu\textrm{s}$ are achieved for the Strang splitting scheme with $dt_\mathrm{1D}=2 \textrm{ or }4\,\upmu\textrm{s}$. The resulting speedups are depicted in Fig.~\ref{fig:SplittingEfficiency}. We normalize the compute times with the compute time of the Godunov scheme with $dt_\mathrm{1D}=0.5\,\upmu\textrm{s}$. For a relative error limit $10^{-3}$ in $V_\mathrm{m}$, one can achieve a speedup of about $7.54$ with the improved time stepping scheme if the accuracy requirement is weakened slightly. If the error constraint cannot be weakened, we still obtain a speedup of $3.89$. Note that, for lower error limits, the speedup achieved with a second-order scheme is even higher due to the higher convergence order.

\subsection{Parallelization -- Domain Partitioning}
\label{sec:domain_decomposition}

The parallelization of the skeletal muscle simulation is based on a domain partitioning approach, i.e., the whole 3D cuboid muscle domain is decomposed into disjoint subdomains (partitions) that are assigned to parallel processes.

\begin{figure}[h!]
  \begin{center}
    \includegraphics[width=0.65\textwidth]{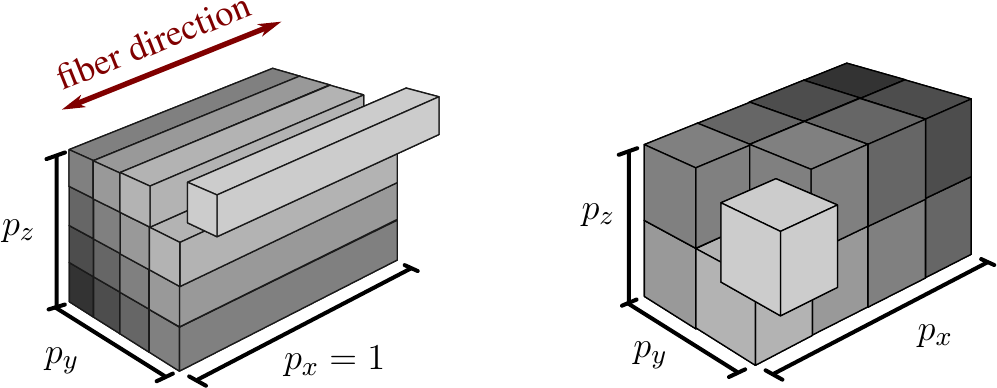}
  \end{center}
  \vspace{-3mm}
  \caption{Visualization of pillar-like (left) and cubic (right) domain decomposition approaches. Both depicted approaches partition the same domain into 16 subdomains with $p_x$, $p_y$, and $p_z$ subdivisions in $x$-,
  $y$-, and $z$-direction, respectively.}
  \label{fig:dd}
\end{figure}

Within this work, we aim to use a parallelization scheme that is able to make use of an arbitrary number of processes. 
Previous work \cite{Heidlauf2013} based their parallelization strategy on physiological characteristics, which leads to pillar-like partitions ensuring that one fiber is never distributed to multiple processors. 
Considering not only the skeletal muscle fibers but the entire model, this induces a significant communication overhead due to the long and stretched partitions. We therefore investigate a further domain partitioning approach with cube-shaped partitions exhibiting sub-domains with minimal surface. 
The two above-mentioned partitioning strategies, \iec the pillar-like and cubic ones, are visualized in Fig. \ref{fig:dd}.

The domain partitioning induces communication due to dependencies of local data on data of neighboring partitions.
These dependencies comprise the following:
\begin{enumerate}
\item \label{observation1} For solving for the propagation of $V_\mathrm{m}$, \iec using an implicit Euler 
 or Crank-Nicolson method (Eqn. \eqref{eqn:CN}) to solve the monodomain equation (Eqn. \eqref{eqn:monodomain}), communication between neighboring processes along a single fiber
is required. The cost for this communication per process grows linearly with the number of fibers contained in the respective 
3D partition. For the existing pillar-like implementation of partitioning, this cost vanishes since fibers are not partitioned into several pieces.
\item \label{observation2} Computing the mucle displacements $\boldsymbol{u}$ of the 3D model (Eqn. \eqref{e:3D_momentum_balance}) involves all processes. This is a result of using a finite element discretization, which inherently requires communication between processes which share common partition boundaries. Thus, these costs are proportional to the surface of the 3D partitions.
\item \label{observation3} Interpolating the muscle displacements $\boldsymbol{u}$ of the 3D muscle mesh to 1D fiber mesh node positions and calculating $l_\mathrm{hs}$, requires ghost layers at the partition boundaries containing one layer of 3D elements.
Note that for the reverse transfer, the accumulation of the activation parameter $\gamma$ from the 0D model at the Gauss points of the elements in the 3D mesh, \iec computing $\bar{\gamma}$, does not involve communication since the process is completely local as all 0D points are contained within the respective 3D element and reside on the same process.
\end{enumerate}

\subsubsection{Generalized and Optimized Domain Partitioning}

Based on the communication dependencies \ref{observation1} and \ref{observation2}, we enhance the original pillar-like domain partitioning in two ways: (i) we allow for an
arbitrary number of processes instead of a fixed number of four processes and (ii) we introduce a new partitioning concept with nearly cubic partitions. 
In contrast to partitioning strategies based on space-filling curves such as \cite{schamberger2005Partitioning}, or graph-partitioning such as \cite{miller1993automatic, zhou2010controlling}, a cuboid partition has the advantage that the interaction of one cuboid partition with others is guaranteed to be planar and bounded by the maximum number of neighboring partitions, \iec $3^3-1=26$. This allows communication with reduced complexity and cost. On the other hand, load balancing may not be optimal because no partition might exist that divides the 3D elements into equally sized sub-domains with a number matching the total number of processes. 

Within this work, we assume that it is possible to create nearly optimal cubic partitions. 
However, we can not completely avoid obtaining sub-domains at the boundary of the computational domain that have less elements than others. 
Given a fixed number of available cores, we maximize the number of employed processes by adapting the number of sub-divisions in each axis direction
corresponding to a factorization of the total number of processes. 
By introducing the additional constraint that each generated partition has to be larger than a specified “atomic” cuboid of elements, we can easily ensure that each process contains only entire fibers (pillar-like partition), a fixed number of fiber subdivisions (cube-like partition), or anything in between. 

In Sec.~\ref{subsec:scaling}, various runtime and scalability studies are presented. In particular, we present scaling results that demonstrate weak scaling for the full muscle model showing the influence of pillar-like or cube-like partitioning and the memory consumption per process of the current implementation in two different experiments; and strong scaling for the full muscle model showing the dependency of the achieved efficiency on the shape of the partitions.

\subsubsection{Optimized Interpolation and Homogenization}

\begin{figure}[h!]
\centering
  \includegraphics[width=0.8\textwidth]{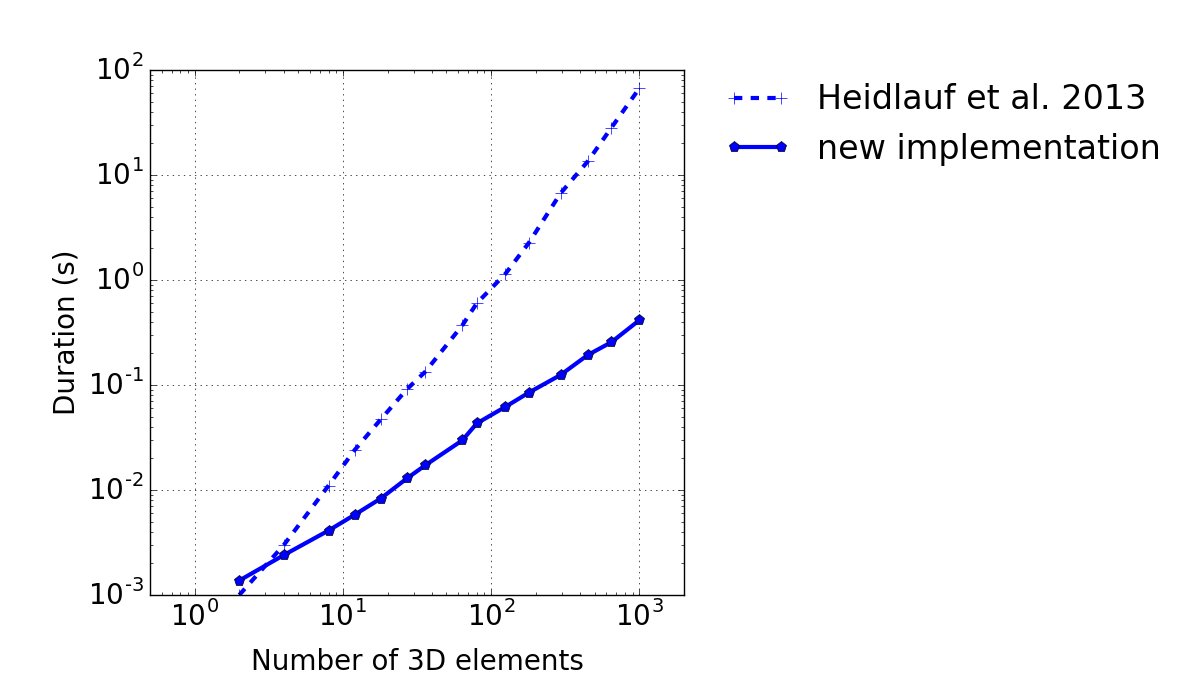}
  \caption{Runtime for the homogenization of values computed on fibers to 3D elements, \iec $\gamma \rightarrow \bar{\gamma}$.}
  \label{fig:cuboid_homogenization}
\end{figure}

Based on the communication dependency \ref{observation3} listed above, the interpolation and homogenization routines need to be optimized. 
Interpolation and homogenization involves information transfer between values at Gauss points of the 3D elements to nodes of the 1D fibers. To allow general domain decomposition, a mapping between the respective 3D and 1D finite elements is necessary. In \cite{Heidlauf2013}, the homogenization was achieved by a naive search over all locally stored fibers. This search was performed for each 3D element. We replace this approach, which exhibits quadratic complexity (in terms of the number of involved elements), by a linear algorithm that uses a correct indexing approach based on the element's topology. Fig.~\ref{fig:cuboid_homogenization} shows that the new implementation achieves substantial runtime reductions.

\subsubsection{Scaling Experiments}\label{subsec:scaling}

All scaling experiments are conducted on HazelHen, the Tier-1 supercomputer at the High Performance Computing Center Stuttgart (HLRS). 
A dual-socket node of this Cray XC40 contains two Intel Haswell E5-2680v3 processors with base frequency of 2.5 GHz, maximum turbo frequency of 3.3 GHz, 12 Cores each and 2 Hyperthreads per Core, leading to a total number of 48 possible threads per core.

\textit{Weak Scaling Measurements -- Experiment \#1.}
For weak scaling, the problem size is increased proportional to the number of processes. Thus, invariants are the number of elements per process and the overall shape of the computational domain. Here, we show weak scaling for both partitioning strategies: partitioning only in $y$- and $z$-direction, i.e., pillar-like partitioning, and cubic partitioning. 
We start with $24$ processes on a single node of HazelHen with an initial partition consisting of $p_x \times p_y \times p_z = 1 \times 6 \times 4 = 24$ subdivisions for both the pillar-like and the cubic partitioning. Each partition contains $e_x \times e_y \times e_z = 2 \times 2 \times 2 = 8$ 3D elements per MPI rank. Further, we ensure that each 3D element contains $2 \times 2$ fibers in $x$-direction with three 1D elements per fiber, i.e., $12$ 1D elements per 3D element. Hence, the initial problem is made up of $24 \times 8 = 192$ elements and $12 \times 8 \times 4 = 384$ fibers.

\begin{figure}[h!]
\centering
  \includegraphics[width=1\textwidth]{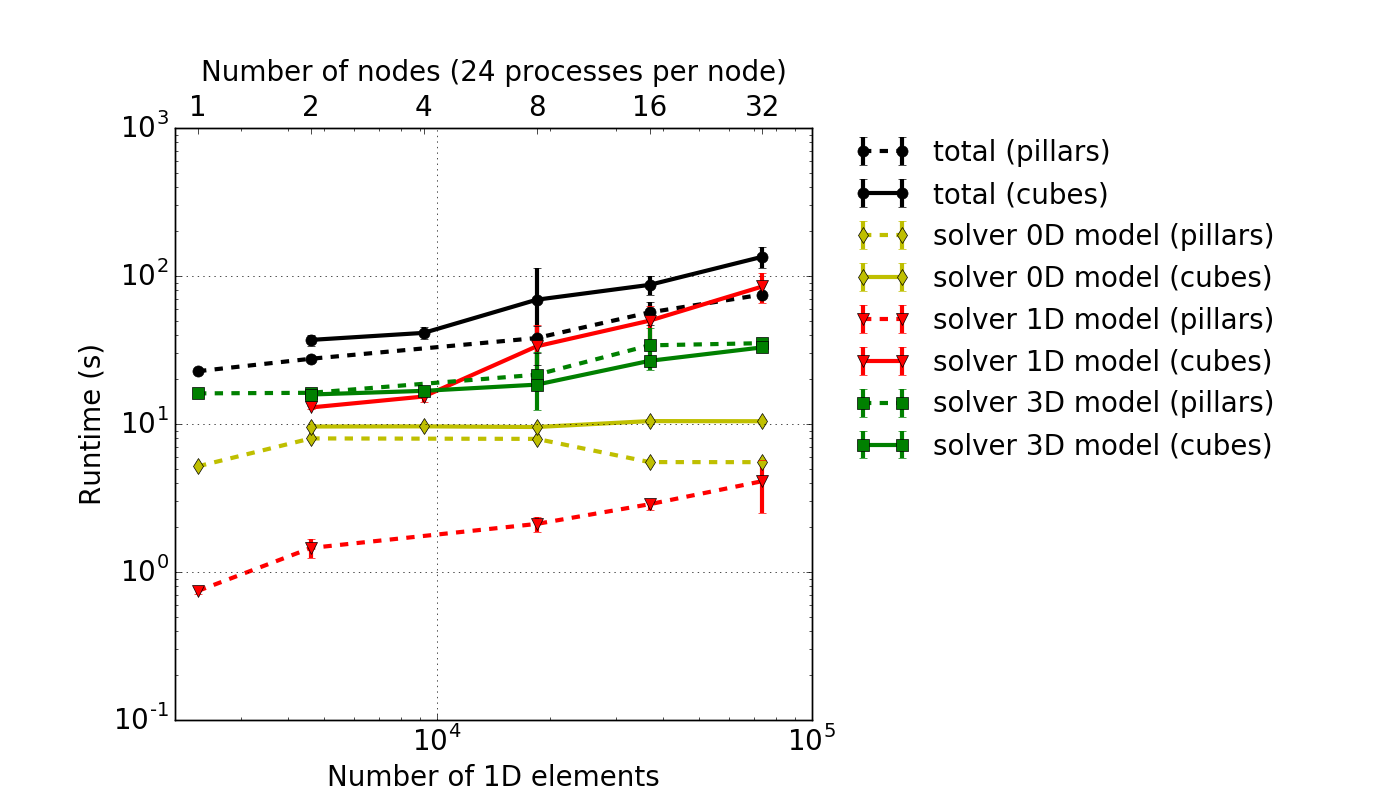}
  \caption{Weak scaling measurements -- experiment \#1: Total runtime as well as individual runtimes (solver for the 0D, 1D and 3D problem) for the cubic partitioning (solid lines) and pillar-like partitions (dashed lines).}
  \label{fig:cuboid_weak_scaling_with_3d}
\end{figure}

In the two conducted series of measurements for the two partitioning strategies, further subdivisions are defined such that the pillar-like or cubic partitioning structure is maintained.
The refinements are obtained by first refining by a factor of $2$ in $x$-direction, in $z$-direction and then in $y$-direction before repeating the process. 
For the cubic partitioning, we fix the number of 3D elements that each MPI rank contains to be $2 \times 2 \times 2$.
For the pillar-like partitioning, the constraint is that each sub-domain spans over all three-dimensional elements in the $x$-direction, whose number varies with increasing problem size. Therefore, the number of elements per MPI rank in $y$- and $z$-direction is halved for each refinement in an alternating way. This way, we double the number of partitions while maintaining the constant number of eight 3D elements per MPI rank. By allocating $24$ processes on the $24$ cores of each node (no hyperthreading), we scale from 1 to 32 nodes, \iec from 24 to 768 cores. Table \ref{tab:PartitionsWeakScaling} provides the details on the partitioning and the number of three-dimensional and one-dimensional elements.
Results are depicted in Fig.~\ref{fig:cuboid_weak_scaling_with_3d}.
\begin{table}[h!]    
	\begin{tabular}{|c|c|r|c|c|c|c|}
	\toprule
	\multicolumn{3}{|l|}{} & \multicolumn{2}{c|}{\textbf{Pillars}} & \multicolumn{2}{c|}{\textbf{Cubes}} \\
	\midrule
	Nodes & 3D Elements & 1D El. & $p_x \times p_y \times p_z$ & $e_x \times e_y \times e_z$ & $p_x \times p_y \times p_z$ & $e_x \times e_y \times e_z$ \\
	& $p_x e_x \times p_y e_y \times p_z e_z$ & & & & & \\
	
	\midrule
	$1$ & $2 \times 12 \times 8$ & 2.304 & $1 \times 6 \times 4$ & $2 \times 2 \times 2$ & $1 \times 6 \times 4$ & $2 \times 2 \times 2$ \\
	$2$ & $4 \times 12 \times 8$ & 4.608 & $1 \times 6 \times 8$ & $4 \times 2 \times 1$ & $2 \times 6 \times 4$ & $2 \times 2 \times 2$\\
	$4$ & $4 \times 12 \times 16$ & 9.216 & $1 \times 12 \times 8$ & $4 \times 1 \times 2$ & $2 \times 6 \times 8$ & $2 \times 2 \times 2$\\
	$8$ & $4 \times 24 \times 16$ & 18.432 & $1 \times 12 \times 16$ & $4 \times 2 \times 1$ & $2 \times 12 \times 8$ & $2 \times 2 \times 2$ \\
	$16$ & $8 \times 24 \times 16$ & 36.864 & $1 \times 24 \times 16$ & $8 \times 1 \times 1$ & $4 \times 12 \times 8$ & $2 \times 2 \times 2$ \\
	$32$ & $8 \times 24 \times 32$ & 73.728 & $1 \times 24 \times 32$ & $8 \times 1 \times 1$ & $4 \times 12 \times 16$ & $2 \times 2 \times 2$ \\
	\bottomrule
	\end{tabular}  
\caption{Weak scaling measurements -- experiment \#1: Problem and partition sizes for 1 to 32 nodes, i.e., $24$ to $768$ cores of HazelHen.}
\label{tab:PartitionsWeakScaling}
\end{table}
The results of Fig.~\ref{fig:cuboid_weak_scaling_with_3d} show that the solver for the 3D model has a slightly higher computational time for the pillar-like partitioning compared to the cubic partitioning. This is expected as the partition boundaries are larger and induce more communication. 
For the 1D problem solver, pillars are better as fibers are not subdivided to multiple cores and no communication is needed. The reduced benefit from a cubic partitioning is due to the fact that time spent on communication is rather dominant compared to the time needed to solve the rather small problem, \egc only $3\,e_x = 6$ 1D elements of a fiber are locally stored in each partition. This should improve as one chooses larger sub-problem sizes, \iec increases the number of nodes per fiber. 

Theoretically, the time needed to solve the 0D problem should not be affected by the domain decomposition. However, due to memory effects, the runtime for a cubic partitioning is slightly higher. Overall, this leads to a higher total computational time for cubic partitioning compared to the pillar-like partitioning. This conclusion is, however, only valid for the chosen scenario and for the relatively low number of cores.
Note that extending this scaling experiments to a larger numbers of cores is currently limited due to memory duplications in the current code. This needs to be first eliminated before conducting further scaling studies.

\begin{figure}[h!]
\centering
  \includegraphics[width=0.9\textwidth]{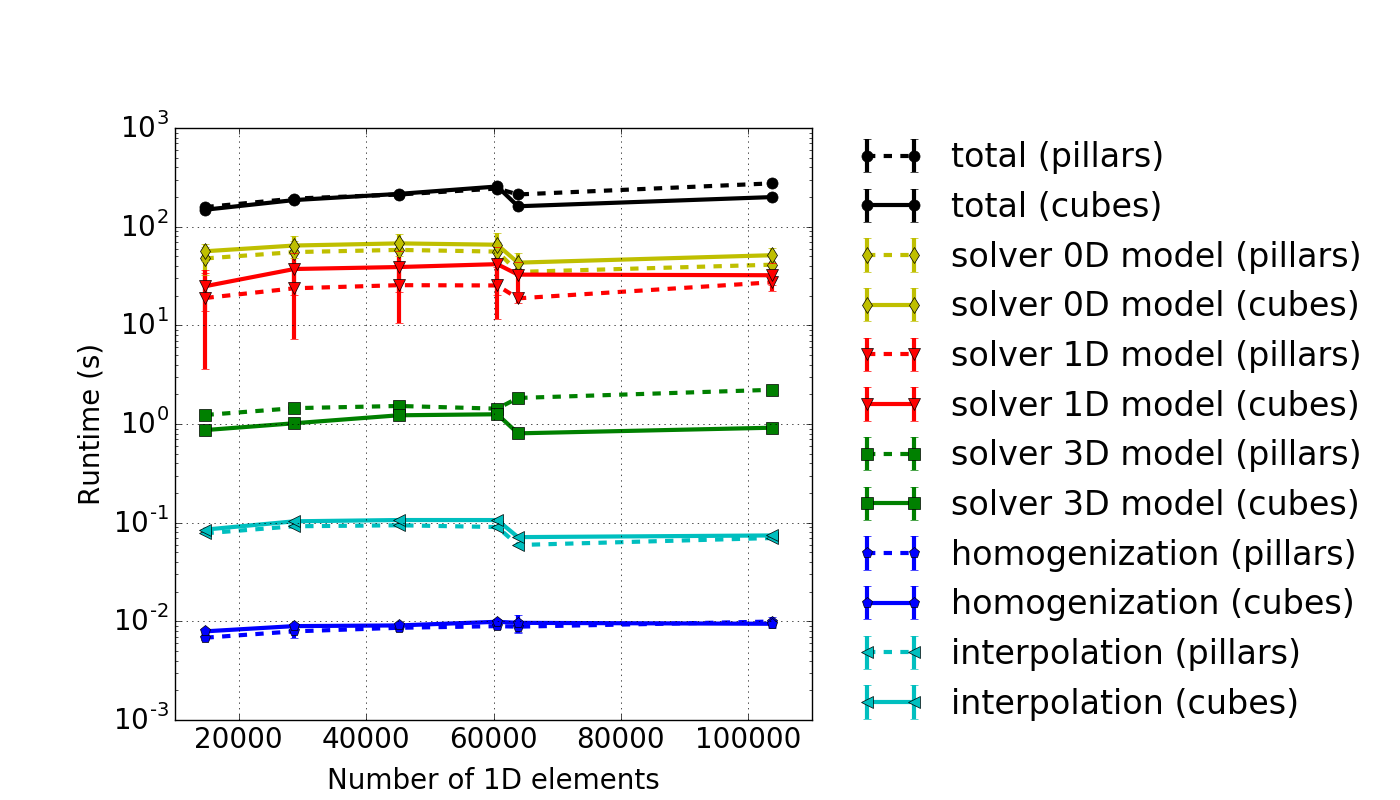}
  \caption{Weak scaling measurements -- experiment \#2: Runtimes for different model components. The results for cubic and pillar-like partitions are depicted by solid and dashed lines, respectively. Different runtime components are encoded in colors, \iec  the total runtime in black, 0D solver in yellow, the 1D solver in red, the 3D solver in green, the interpolation in light blue and the homogenization in dark blue.} 
    \label{fig:cuboid_weak_scaling_multi_equal_p}
 \end{figure}

\textit{Weak Scaling Measurements -- Experiment \#2.}
While the somewhat artificial setting in experiment \#1 yields perfect pillar-like or cubic partitions, experiment \#2 addresses a more realistic setup, where we increase the number of processes more smoothly, i.e., by less than a factor of two in each step. With this, it not possible any more to choose perfect cubic or pillar-like partitions. Thus, we identify reasonable parameters by solving an optimization problem that trades based on process counts and targeted sub-domain shape. Note that the combination of number of processes and number of elements leads to partitions at the boundary of the computational domain that potentially have less elements than interior partitions. Compared to the previous example, the number of 3D elements per process is only approximately constant with the pillar-like
partitions getting closer to constant size than the cubic ones. The numbers of processes and dimensions of the computational domain are listed in Table ~\ref{tab:WeakScalingExperiments2}. Fig.~\ref{fig:cuboid_weak_scaling_multi_equal_p} presents the runtime results. \\

\begin{table}[h!]    
	\begin{tabular}{|l|c|r|c|c|c|c|}
	\toprule
	\multicolumn{3}{|l|}{} & \multicolumn{2}{c|}{\textbf{Pillars}} & \multicolumn{2}{c|}{\textbf{Cubes}} \\
	\midrule
	Nodes, & 3D Elements                        & 1D El. & $p_x \times p_y \times p_z$ & $e_x \times e_y \times e_z$ & $p_x \times p_y \times p_z$ & $e_x \times e_y \times e_z$ \\
	Cores  & $p_x e_x \times p_y e_y \times p_z e_z$  &                   & & & \\
	\midrule
	1, 24 &$16 \times 11 \times 7$   & 14.784 & $1 \times  6 \times 4$ & $16 \times 2 \times 2$ & $4 	\times 3 \times 2$ & $4 \times 4 \times 4$\\
	2, 40 &$18 \times 19 \times 7$   & 28.728 & $1 \times 10 \times 4$ & $18 \times 2 \times 2$ & $4 	\times 5 \times 2$ & $5 \times 4 \times 4$\\
	3, 60 &$18 \times 19 \times 11$  & 45.144 & $1 \times 10 \times 6$ & $18 \times 2 \times 2$ & $4 	\times 5 \times 3$ & $5 \times 4 \times 4$\\
	4, 84 &$17 \times 27 \times 11$  & 60.588 & $1 \times 14 \times 6$ & $17 \times 2 \times 2$ & $4 	\times 7 \times 3$ & $5 \times 4 \times 4$\\
	6, 140 &$38 \times 20 \times 7$  & 63.840 & $1 \times 20 \times 7$ & $38 \times 1 \times 1$ & $10	\times 7 \times 2$ & $4 \times 3 \times 4$\\
	8, 192 &$45 \times 16 \times 12$ & 103.680 & $1 \times 16 \times 12$& $45 \times 1 \times 1$ & $12	\times 4 \times 4$ & $4 \times 4 \times 3$\\
 \bottomrule
\end{tabular}
\caption{Weak scaling measurements -- experiment \#2: Number of elements, number of fibers and partition sizes.}
\label{tab:WeakScalingExperiments2}
\end{table}
As already discussed above, the ODE solver for the 0D-problem (yellow line) requires the majority of the runtime. 
This is followed by the solution times for the 1D (red line) and 3D (green line) sub-problems. 
The blue lines depict the duration of the interpolation and homogenization between the node positions of the 1D fibers and the 3D mesh. 
It can be seen that the computational times stay nearly constant for increasing problem size. As in the previous experiment, the 3D solver performs better for cubic partitioning whereas the 1D solver is faster for pillar-like partitions. In this scenario, the cubic partitioning in total slightly outperforms the pillar-like partitioning, as expected.
		
\begin{figure}[h!]
\centering
    \includegraphics[width=1.0\textwidth]{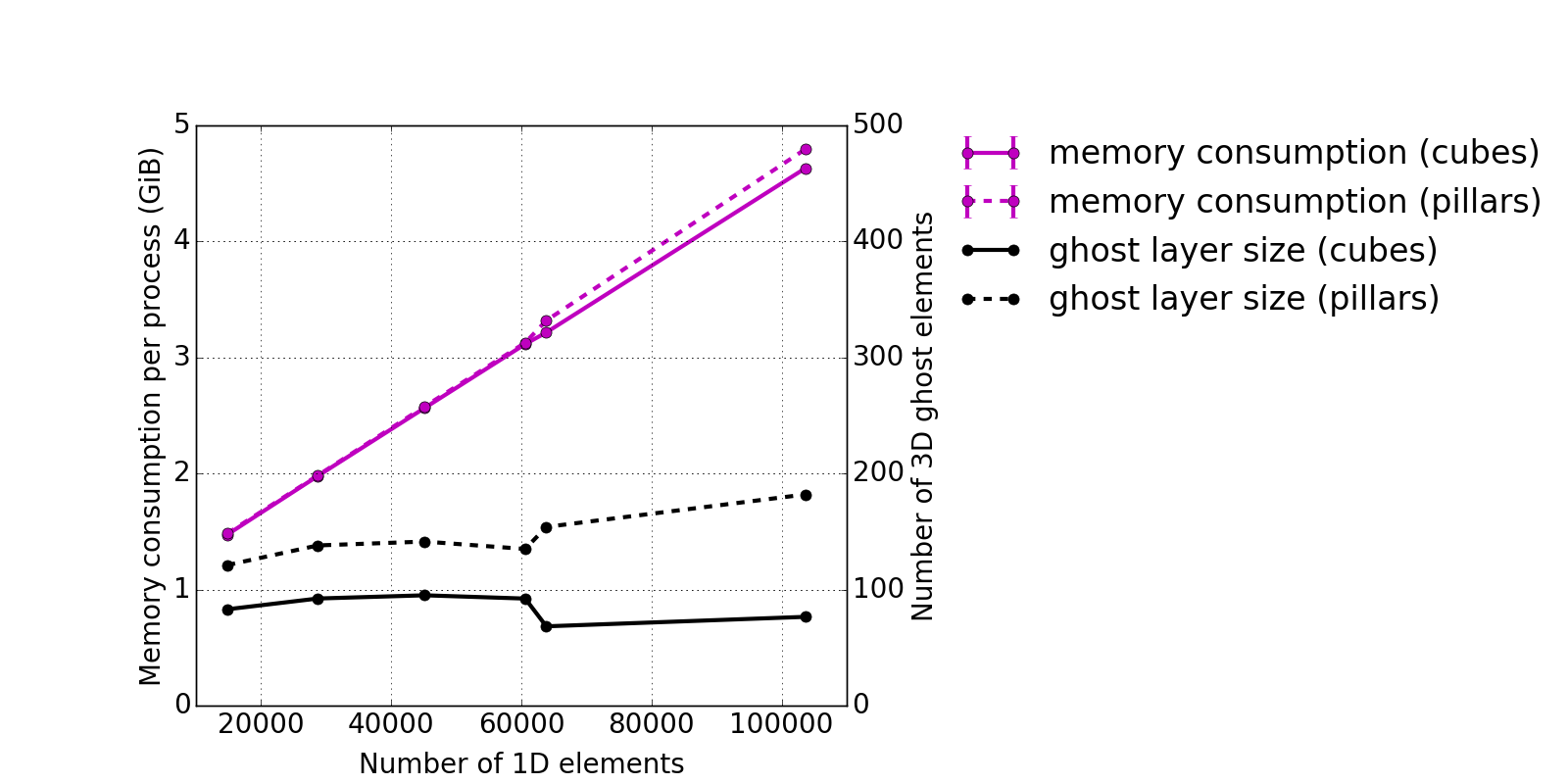}
    \caption{Weak scaling measurements -- experiment \#2: Total memory consumption per process at the end of the runtime. The total memory consumption is depicted in magenta and the average number of 3D ghost layer elements per process in black. Again, the solid lines represent cubic partitioning and the dashed line piller-like partitioning.}
    \label{fig:cuboid_weak_scaling_multi_equal_p_memory}
\end{figure}

As before, the memory consumption appears to be a weakness. Therefore, additional tests investigating the memory consumption per process at the end of the runtime were carried out.
The memory consumption for the presented scenario is plotted in Fig.~\ref{fig:cuboid_weak_scaling_multi_equal_p_memory} with respect to the overall number of 1D elements. Also the average number of ghost layer elemts per process is depicted. Ghost layer elements are copies of elements adjacent to the partition of a process, i.e., belong to the subdomain of a neighbouring process. They are used as data buffers for communication. 

We observe that 
the average number of ghost elements per process for the 3D problem is higher for pillar-like partitions (dashed black line) than for the cubic partitions (solid black line).
A sharp increase of memory consumption (magenta lines) is observed independent of the partitioning scheme. This is due to duplications of global data on each process which will be eliminated in a next optimization step. 
Compared to this effect, the difference between the number of ghost elements needed for the two partitioning strategies is negligible.

\begin{figure}[h]
  \begin{center}
    \includegraphics[width=0.9\textwidth]{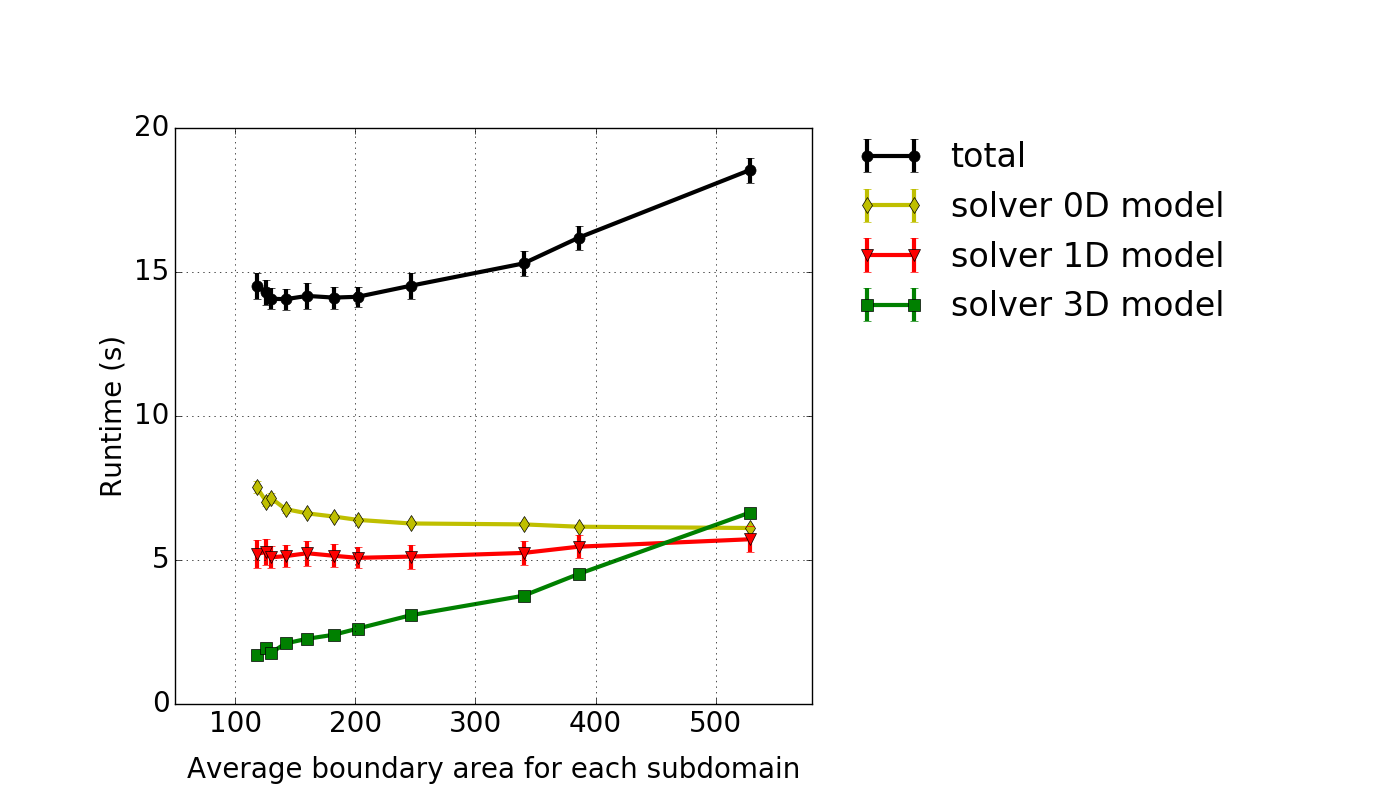}
  \end{center}
  \caption{Dependency between runtime and partition shape -- experiment \#3: Runtime in dependence on the average boundary area of the partitions. We show the accumulated total computational time and the runtimes of the sub-problems as in the previous studies.}
	\label{fig:cuboid_subdivision_runtime_12_12}
\end{figure}

\textit{Dependency Between Runtime and Partition Shape -- Experiment \#3.}
In our third scaling test, the dependency of the solver of the 3D continuum-mechanical problem on the partitioning strategy is investigated. We analyse how different domain decomposition approaches, in particular others than the previously discussed pillar-like and cubic partitioning schemes, affect the runtime. A test case with $144\times 12\times 12$ three-dimensional elements is considered. Otherwise, the setup is 
as described in Sec.~\ref{sec:test_scenario}.
To reduce the contributions of the 0D/1D sub-problem and focus on the performance of the 3D components, we include in each 3D element only two 1D fiber elements.  The domain is decomposed into a constant number of $144$ partitions by axis-aligned cutting planes in all possible ways.
To distinguish between the different partitioning variants, we compute the average boundary surface area between the partitions for each variant and relate this to runtime. The results are presented in Fig.~\ref{fig:cuboid_subdivision_runtime_12_12}. 
The smallest average surface area between the partitions, which corresponds to the first data point in Fig.~\ref{fig:cuboid_subdivision_runtime_12_12}, is obtained for a partitioning with $144$ partitions with $4\times 6\times 6$ elements each. 
The highest average surface area between the partitions, which is the last data point within Fig.~\ref{fig:cuboid_subdivision_runtime_12_12}, is obtained for $144$ partitions with $1 \times 12 \times 12$ elements each. 
All experiments are run on 12 nodes of Hazel Hen with 12 processes per node. 
It can be seen that only the time needed to solve the 3D continuum-mechanical problem increases monotonically with respect to the average surface area between the partitions, \iec depends on the partitions' shape. This is expected. Further, the runtime ratio of the 3D solver between the partitioning with the smallest and largest average surface area is $1 : 4.3$.

\subsection{Optimizing Input/Output -- File Format}
\label{sec:file_format}
Optimizing HPC simulations also includes the consideration of efficient input and output (IO) file formats, \egc writing data in parallel to reduce communication, and binary codecs to reduce the size of a data set.
Currently, OpenCMISS appeals to an EX-file format\footnote{\url{http://opencmiss.org/documentation/data_format/ex_file_format.html}, Accessed: 2017-09-30}, which writes header information and data as plain text.
Headers and data are stored in the same file, and one file is generated for each time step.
In general, writing text-based data has very low read and write performance, and very high storage requirements compared to binary encoded data.
To successfully run simulations on large systems, the data IO should ideally not cause scaling issues.
Therefore, we propose in this work a new, efficient file format\footnote{\url{https://github.com/UniStuttgart-VISUS/ngpf}, Accessed: 2017-09-30}.

\begin{figure}
\centering
\includegraphics[width=0.75\linewidth]{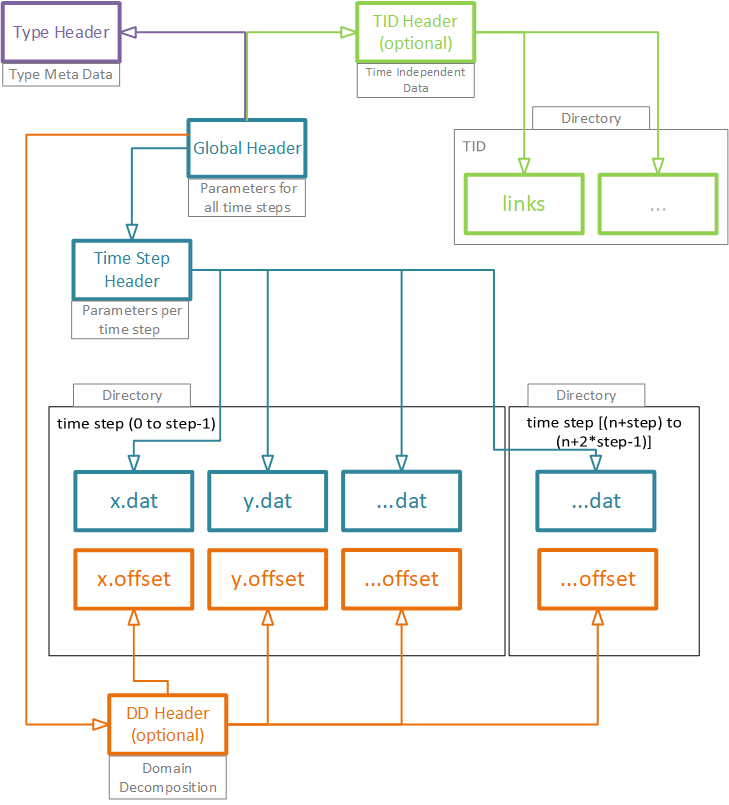}
\caption{Header and data scheme for the file format.
The blue boxes illustrate the mandatory, basic data.
Attribute data (x.dat, y.dat) are written to individual files.
This can be extended by several optional data (violet, green, orange).
The violet type header can provide additional meta data such as a charge, name, or structural information.
The green time-independent header can be used to write additional data that do not change during a simulation, but applies for each simulated component, e.g., links between mesh vertices.
The orange domain decomposition extension stores information about the decomposition, e.g., data written per node or partition (cells, cell compositions).
}
\label{fig:file_format_schematic}
\end{figure}

The new data structure includes custom binary codecs, such as the high-compression algorithm ZFP~\cite{Lindstrom2014}, and stores meta data in separate header files using the human-readable JSON object notation.
Fig.~\ref{fig:file_format_schematic} gives an overview of the data layout, where each colored box denotes a separate file.
The main feature of the new file format is the full support of domain decomposition (DD) with parallel file IO.
Every node processes its own set of data (e.\,g., time integration or data encoding) and claims the corresponding memory region in the data files to write the encoded data stream.

The new file format has three optional features.
It can write time independent data (TID, cf. Fig.~\ref{fig:file_format_schematic}) that are written per attribute data, e.\,g., positions of Gaussian points in the 3D mesh in the reference configuration. 
A further feature is the \textit{"Type Header"} that writes type-specific quantities, which can contain simple attributes or a complex structural layout.
With the third feature, the DD meta data (e.\,g., load in each node), we are able to analyze how each node has performed during the simulation.
These meta data can answer questions like ``which muscle fiber was simulated on which node'', or ``was this node at full capacity at a specific point in time''.
These data can be used to optimize the load balancing within the simulation in future work.

\subsection{Visualization}

Efficient (parallel) IO-file formats are also the basis for the visualization of large data sets. 
The current visualization framework within the OpenCMISS software initative is OpenCMISS-Zinc.
This framework already offers a range of visualization techniques for muscle fiber data, e.\,g., a convex hull calculation to construct a mesh geometry from point cloud data.
However, Zinc uses outdated rendering methods, resulting in insufficient performance for large-scale data.
Furthermore, it does not offer a suitable platform for fast visualization prototyping, distributed rendering, or advanced visualization techniques, which would be required for developing new muscle fiber visualizations.
MegaMol \cite{Grottel2015} fulfills these criteria and offers additional functionality and features that are valuable for this project.
One example for an additional feature is the provided infrastructure for brushing and linking that allows for interactive visual analytics.
MegaMol also offers a built-in headless mode and a remote control interface that is crucial for the HPC and in-situ rendering planned for the future in order to support the visual analysis of large-scale muscle simulations.
In-situ visualization is an alternative approach to traditional post-hoc data processing.
It processes and visualizes data while the simulation produces them. 
Consequently, writing raw data to disk can be avoided completely.
In-situ visualization is very much an area of active research, and thus, there are many different approaches and no standard approach has been established (see, e.\,g., \cite{childs2012situ}).

\begin{figure}[htbp]
  \centering
	\includegraphics[width=1.0\linewidth]{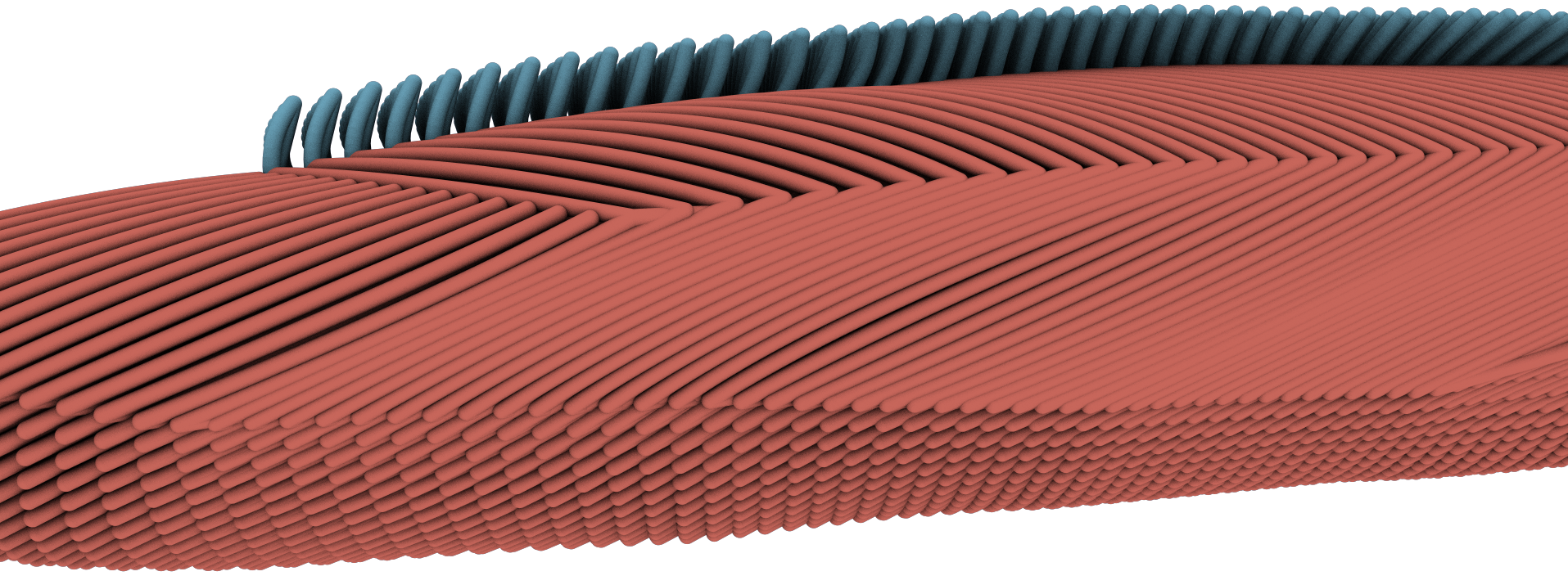}
	\caption{The discretised 1D muscle fibers are rendered as continuous tubular streamlines to show the characteristics and implicit geometry of the individual strands.
	The color coding can be used to show the distribution of parameter values along the fibers.}
	\label{fig:muscle_lines}
\end{figure}

MegaMol allows GPU and CPU rendering via a thin abstraction layer, where the GPU rendering uses the OpenGL API, whereas the CPU rendering is based on the ray tracing engine OSPRay \cite{Wald2017}.
In particular the CPU-based rendering enables image synthesis on any HPC cluster, regardless of the availability of dedicated GPUs.
It offers advanced rendering and shading methods (e.\,g., path tracing and ambient occlusion) that enhance the perception of depth and run mostly interactive even on single workstations.
Like the simulation software OpenCMISS, MegaMol is currently not optimized for HPC usage, but it provides a solid base and infrastructure and is already capabale of rendering the discretized muscle fibers as continuous geometry (cf.\ Fig.~\ref{fig:muscle_lines}).

While analyzing and optimizing existing code for HPC infrastructures are best performed with test cases, in which the geometry has a minimal influence, \egc the cuboid muscle test case as introduced in Sec.~\ref{sec:test_scenario}, it is advantageous to demonstrate the visualization capabilities on complex geometries. 
Therefore, data from previous simulations (here data of the Tibialis Anterior as obtained within \cite{Heidlauf2013}) are used to showcase visualization capabilities. 
The image was rendered with MegaMol \cite{Grottel2015} using the CPU ray tracing engine OSPRay \cite{Wald2017}.

\section{Conclusion and Outlook}

Using models to gain new insights into the complex physiological or anatomical mechanisms of biological tissue, or to better interpret and understand experimentally measured data, requires accurate and detailed models of the underlying mechanisms. 
Depending on the application or the system to be modeled, the required detail can lead very quickly to highly complex and computationally extremely demanding models. 
Most software packages are, however, often designed to build up computational models for a variety of complex biomechanical systems, like in the case of OpenCMISS for simulating the chemo-electromechanical behavior of skeletal muscles after neural recruitment, the mechanics of the heart, the functioning of the lung, etc.
Such software packages might already run within a parallel computing environment, but are not optimized to run large-scale simulations on large-scale systems such as HazelHen, the Tier-1 system in Stuttgart.
Before being able to exploit the full capabilities of supercomputers, software packages such as OpenCMISS have to be analyzed and optimized to achieve good scaling properties -- ideally perfect scaling meaning that the simulation of a twice as large problem on twice as many nodes/cores requires the same runtime as the original setup. 
 
Within this paper, we have demonstrated that the chemo-electromechanical multi-scale skeletal muscle model as introduced in Sec.~\ref{sec:model} and implemented in OpenCMISS is capable of running significantly sized model setups in a parallel compute environment. We have simulated the deformation of a skeletal muscle in which 34,560 randomly activated fibers are discretized with 103,680 1D elements.
Further, by utilizing a standard test case, we have been able to show good weak scaling properties for a small number of compute nodes. 
For the partitioning of the domain, two different approaches have been considered: a pillar-like partition, in which each embedded 1D skeletal muscle fiber is kept within a single partition, and a cubic partitioning, in which individual fibers can be distributed over different partitions. 
It has been observed that mainly the solution times of the 3D and the 1D solver depend on the domain partitioning. For the test cases, the 1D solver profits from pillar-like partitions, while the 3D solver exhibited lower runtimes for cubic partitions. 
The analysis has revealed that for compute-intense scenarios with a high numbers of 1D elements (in relation to the computational work needed to solve the 3D continuum-mechanical model), cubic partitions yield a better overall performance. 
The scaling of the 1D solver with increasing problem size has been found to be linear and, hence, promises good behavior when the overall problem size will be further increased in the future. 
Furthermore, the text based EX-file format has been identified as not appropriate to run on an HPC system. Therefore, a new file format is being developed that offers high-compression algorithms and performs file IO in parallel.
However, before utilizing significant portions of HazelHen, further aspects concerning the model, algorithms, implementation, and visualization need to be considered. 

From a modeling point of view, we would like to use even more complicated chemo-electromechanical model that include, for example, the mechanical behavior of titin \cite{Heidlauf2016, Heidlauf2017}, and include further important biophysical details such as metabolism, a biophysical recruitment model \cite{Heidlauf2013b}, and a feedback mechanism from the spindles and the golgi-tendon organs to the neuromuscular system. Moreover, simulating and visualizing the surface EMG is needed to further test motor unit decomposition algorithms. 
From an algorithmic point of view, such model enhancements also always require novel or custom-tailored efficient numerical schemes. 
From an implementational point of view, the scaling analysis also revealed a memory consumption issue. 
The amount of memory per process rises proportional to the total problem size. The cause for this has been identified as the suboptimal implementation of a global-to-local node numbering mapping within OpenCMISS, which, however, can be fixed. 
The memory consumption currently prevents simulation of larger problem sizes. 

Once the modeling, algorithmic, implementational aspects have been solved and large-scale simulations have been set up, \egc a single chemo-electromechanical skeletal muscle model with a realistic number of fibers (\egc about 300.000) of realistic length, the results need to be visualized and analyzed. 
To do so, we are planning to extend MegaMol with novel, comprehensive visualizations that allow users to explore the complex behavior of muscle fiber simulation data.
For example, the deformation due to the contraction of the muscle is currently barely visible for different time steps, whereas to compare the contraction of different runs or different time steps, a clear visual representation is necessary.
Furthermore, visualization can help to gain new insights into the functioning of our neuro-muscular system by comparing the simulated surface EMG 
of a muscle with experimental data obtained via non-invasive and clinically available diagnostic tools. 

After we achieved all the model enhancements and have overcome the computational and implementational challenges, these simulations can be utilized to set up in silico benchmark tests for complex muscles and recruitment patterns to improve existing data mining algorithms, e.g., the motor unit decomposition algorithms, and provide, at the same time, a new tool to investigate the interplay of the underlying complex and coupled mechanisms leading from neural stimulation to force generation.

\section*{Conflict of Interest Statement}

The authors declare that the research was conducted in the absence of any commercial or financial relationships that could be construed as a potential conflict of interest.

\section*{Author Contributions}
All authors have equally contributed to the conception and design of the work, data analysis and interpretation, drafting of the article, and critical revision of the article. Hence, the author list appears in alphabetical order. In addition NE, TK, AK, BM and TR have  conducted the simulations and summarized their results. All authors fully approve the content of this work.

\section*{Funding}
This research was funded by the Baden-W\"urttemberg Stiftung as part of the DiHu project of the High Performance Computing II program, the Deutsche Forschungsgemeinschaft (DFG) as part of the International Graduate Research Group on "Soft Tissue Robotics – Simulation-Driven Concepts and Design for Control and Automation for Robotic Devices Interacting with Soft Tissues" (GRK 2198/1) and as part of the Cluster of Excellence for Simulation Technology (EXC 310/1).

\bibliographystyle{abbrv}
\bibliography{bib_for_this_pub}

\end{document}